# The ontogeny of discourse structure mimics the development of literature

Natália B. Mota [1*], Sylvia Pinheiro [1*], Mariano Sigman [2], Diego Fernández Slezak [3,4], Guillermo Cecchi [5], Mauro Copelli [6#], Sidarta Ribeiro [1#]

* Equal contribution
# Corresponding authors

1 - Instituto do Cérebro, Universidade Federal do Rio Grande do Norte, Natal, Brazil.
2- Universidad Torcuato Di Tella, CONICET, Buenos Aires, Argentina.
3 - Departamento de Computación, Facultad de Ciencias Exactas y Naturales, Universidad de Buenos Aires, Buenos Aires, Argentina.
4 - Instituto de Investigación en Ciencias de la Computación, CONICET, Universidad de Buenos Aires, Buenos Aires, Argentina.
5 - Computational Biology Center – Neuroscience, IBM T.J. Watson Research Center, Yorktown Heights, USA.
6 - Departamento de Física, Universidade Federal de Pernambuco, Recife, Brazil.

**Acknowledgements:**
Work supported by UFRN, Conselho Nacional de Desenvolvimento Científico e Tecnológico (CNPq), grants Universal 480053/2013-8 and Research Productivity 306604/2012-4 and 310712/2014-9; Coordenação de Aperfeiçoamento de Pessoal de Nível Superior (CAPES) Projeto ACERTA; Fundação de Amparo à Ciência e Tecnologia do Estado de Pernambuco (FACEPE); Center for Neuromathematics of the São Paulo Research Foundation FAPESP (grant # 2013/07699-0), Boehringer-Ingelheim International GmbH (contract # 270561). We thank M Posner, S Dehaene, S Bunge, CJ Cela Conde, S Lipina, D Araujo, C Queiroz, J Sitt, JV Lisboa, A Cabana, J Queiroz, J Luban, LF Tófoli, and A Guerreiro for insightful discussions; M Laub and JE Agualusa for source material, PPC Maia for IT support, D Koshiyama for bibliographic support; and Instituto Metrópole Digital UFRN for cloud usage.

**Abstract**

Discourse varies with age, education, psychiatric state and historical epoch, but the ontogenetic and cultural dynamics of discourse structure remain to be quantitatively characterized. To this end we investigated word graphs obtained from verbal reports of 200 subjects ages 2-58, and 676 literary texts spanning ~5,000 years. In healthy subjects, lexical diversity, graph size, and long-range recurrence departed from initial near-random levels through a monotonic asymptotic increase across ages, while short-range recurrence showed a corresponding decrease. These changes were explained by education and suggest a hierarchical development of discourse structure: short-range recurrence and lexical diversity stabilize after elementary school, but graph size and long-range recurrence only stabilize after high school. This gradual maturation was blurred in psychotic subjects, who maintained in adulthood a near-random structure. In literature, monotonic asymptotic changes over time were remarkable: While lexical diversity, long-range recurrence and graph size increased away from near-randomness, short-range recurrence declined, from above to below random levels. Bronze Age texts are structurally similar to childish or psychotic discourses, but subsequent texts converge abruptly to the healthy adult pattern around the onset of the Axial Age (800-200 BC), a period of pivotal cultural change. Thus, individually as well as historically, discourse maturation increases the range of word recurrence away from randomness.

Culture shapes the organization of discourse in ontogeny as in history. At the individual level, language begins to be learned within weeks of birth if not earlier [1,2] but its full development takes many years of formal and informal education [3,4]. At the historical level, the schooling of readers that become writers led to the gradual development of literature. Since the *edubas* of Sumer, schools are organizations specialized in using the scaffolding of biological maturation to train declarative and procedural skills such as reading and writing, firmly grounded on the progressive expansion of memory capacity, memory retrieval, coordination, brain area recycling and symbolic repertoire [5-8]. While phonological perception and production are typically mastered within the initial years of life, vocabulary, syntax and grammar continue to mature into high school through a combination of cognitive development and education that is accelerated by alphabetization but undergoes an extended period of subsequent refinement [3,4,9-11].

In 1-2% of the population, however, discourse deteriorates during adolescence instead of improving [12,13], despite schooling and in parallel with the first surfacing of psychotic symptoms [14-16]. The progressive mental perturbations that characterize schizophrenia typically appear between adolescence and early adulthood. The contrast between healthy and psychotic development before adolescence is blurred, because children are normally more prone to confabulation than adults [17], and often engage in private speech that includes dialogues with imaginary friends [18]. Indeed, a reliable diagnosis of psychosis before the middle infancy is effectively precluded by the fact that healthy children under ~7 years old normally display illogical thinking and loosening of associations [19].

Two general hypotheses arise in this context. First, if psychosis represents the lingering of immature mental functioning [17-19], the disorganization of language that results from psychosis may follow the reverse path of normal language development. Second, given that the semantic analysis of ancient texts has pointed to psychosis as an early trait of civilization [20-22], psychosis may represent a trace of immature human language at both the ontogenetic and historical levels.

Understanding how discourse develops in time poses a significant mathematical challenge, because the lexicon is a high-dimensional object [23,24]. Since our hypotheses set predictions on the organization of words, the most natural way to examine them in a quantitative manner is to measure graph attributes, which allow for structural network characterization [25], and account for the global organization of the lexicon [26-28]. Psychotic discourse is characterized by reduced vocabulary, short-range repetitions of word sequences, a reduction in the degree to which a word is related to other words (which constrains the emergence of long-range themes), and a decrease in the global extent of the word network employed [12,14,15,29,30]. Each of these aspects corresponds to a specific property in a graph made of words, respectively 1) lexical (node) diversity, 2) short-range recurrence, 3) long-range recurrence, and 4) graph size. These properties successfully grasp disorganized language in psychotic adults [29,30] and language organization during the alphabetization of healthy children [31]. Hence, graph analysis provides natural metrics to establish the distance between healthy and psychotic adults - investigated here as proxies of organized and disorganized discourse - to then test whether verbal reports from healthy children move along this dimension as they mature. For that, interviews from 135 healthy subjects and 65 psychotic patients 2 to 58 years old were recorded and analyzed (**Suppl. Table 1, Suppl. Material 1**).

Importantly, the same metrics can be used to assess the structural dynamics of the cultural record. While written text and transcribed speech differ in many ways, our assumption is that they are structurally comparable. To this end, we initially analyzed 448 representative literary texts spanning 5,500 years **(Suppl. Table 2, Suppl. Material 2)**, comprising the following Afro-Eurasian traditions: Syro-Mesopotamian (N=62), Egyptian (N=49), Hinduist (N=37), Persian (N=19), Judeo-Christian (N=76), Greek-Roman (N=134), Medieval (n=20), Modern (n=20) and Contemporary (n=31).

Previous results [29-31] lead us to predict that as healthy subjects age and undergo schooling, their memory reports should progressively increase in lexical diversity (number of nodes - N), long-range recurrence (largest strongly connected component - LSC) and graph size (average shortest path - ASP). On the other hand, short-range recurrence (repeated edges - RE) should gradually decrease (**Fig. 1A**). Reports from psychotic subjects should not show the same dynamics, i.e. we hypothesize that the same 4 graph attributes will be less correlated with age or years of education, remaining similar to those of healthy children's reports. Finally, we expect the dynamics of graph attributes across the historical record to resemble ontogenetic changes in healthy subjects.

For each dataset, we measured the 4 graph attributes of interest N, LSC, ASP and RE, controlling for differences in total number of words per report by averaging across moving windows of 30 words with 50% of overlap (**Fig. 1B**), as detailed in [30] and **Suppl. Material 3**. The evolution of each attribute was modeled as an exponential fit to represent their accelerated initial development followed by a saturation process of slow progress, with $f(t) = c+(a-c)(1-\exp(-t/\tau))$; where $a$ is the asymptotic graph attribute value, $c$ is the initial graph attribute value observed, and $\tau$ is characteristic time to reach saturation (**Suppl. Material 4**). This fit to exponentials allows us to identify dynamic properties of each attribute and hence examine in a quantitative manner whether the historical development of literary structure mimics the ontogenetic dynamics of verbal discourse. It also sets the stage for specific predictions.

At the ontogenetic level, the saturation onset should either precede or coincide with adolescence, when it becomes possible for the first time to clinically identify the losses produced by psychosis [19]. Furthermore, if discourse in healthy children shifts through development from disorganized to organized, but remains largely disorganized in psychotic subjects, we expect initial and asymptotic graph attribute values to be quite different in the former, but not in the latter, i.e. $|a\text{-}c|$ should be greater in healthy subjects than in psychotic patients. Furthermore, healthy subjects should show $a > c$ for N, ASP and LSC, but $c > a$ for RE.

For several reasons, precise predictions for cultural development are harder to make. The mathematical analysis of ancient texts is inherently impacted by a plethora of confounds, such as imprecise dating, variable physical support, multiple authorship and versions, editing, censorship, standardization, translation, access to few, production by fewer, distinct degrees of versification and fictionalization, stylistic, aesthetic and philosophical differences of both authors and translators [21]. A distinctive limitation is the fact that the transition from orality to literacy can only be timed by approximation, with reference to the earliest texts available (~3,000 AC, **Suppl. Material 5**). Furthermore, the historical evolution of narrative complexity was surely shaped by different schools of literary criticism, since writing at any given time is informed by knowledge and criticism of previous writing forms [32]. The investigation of discourse structure across such different scales of analysis, involving both biological and cultural phenomena, must have categorical limitations that at some point turn potential homology into mere metaphor [33]. Due to their inherently different nature, spontaneous speech and literature, albeit possibly sharing mechanisms for the accumulation of complexity over time, are also expected to differ in many ways. Notwithstanding all these caveats, we expect historical development to overall resemble healthy ontogenetic dynamics, and thus $a\text{-}c$ should be positive for N, ASP and LSC but negative for RE. We also expect the characteristic times of the structural development of literature to either precede or coincide with the Axial Age (800-200 BC, **Suppl. Material 5B**), a period in ancient history marked by a philosophical, artistic, political, legal, economic and educational boom in Afro-Eurasia [34-41].

### Ontogenetic dynamics of discourse structure

The 4 graph attributes differed as predicted between healthy subjects below and above 12 years of age, indicating a change towards more organized discourse (**Fig. 1C,** light and dark blue columns). Also as expected, psychotic subjects produced reports that structurally resembled the disorganized pattern seen in healthy subjects with less than 12 years of age (**Fig. 1C,** light blue and red columns). Importantly, both groups yielded measurements equivalent to those of Pre-Axial literature, while Post-Axial literature structurally resembled reports from healthy adults (**Fig. 1C,** white and black columns**; Suppl. Table 3**).

Representative graphs illustrate the marked structural differences between healthy children and adults, not present in psychotic subjects (**Fig. 2A**). In support of our hypotheses, 3 attributes of interest (N, LSC, ASP) showed significant positive correlations with both age and education in healthy subjects (**Suppl. Tables 4 and 5**). The short-range recurrence attribute RE, which in healthy children is negatively correlated with Intelligence Quotient and Theory of Mind scores [31], showed a significant negative correlation with education but not with age in healthy subjects (**Suppl. Tables 4 and 5**). In striking agreement with our prediction that psychotic language remains in a disorganized stage, none of the graph attributes changed significantly either with age or with education among psychotics (**Suppl. Tables 4 and 5**). A multiple linear correlation (**Suppl. Table 6)** confirmed the predominance of education over age in healthy subjects (compare **Suppl. Tables 4** and **5**).

To further characterize these changes, graph attribute values were binned in years of education, and fit with an exponential model weighted for the standard error of the mean (**Suppl. Material 4**). Graph attributes obtained from healthy subjects adjusted very well to the model (**Fig. 2B-E, blue panels)**, with an education-related exponentially saturating increase in lexical diversity (**Fig. 2B**), and a corresponding decrease in short-range recurrence (**Fig. 2C**). Long-range recurrence (**Fig. 2D**) and graph size (**Fig. 2E**) showed a much slower saturating increase. In agreement with our hypothesis that the organization of psychotic discourse changes less through time, the graph parameters obtained from the recordings of psychotic subjects adjusted poorly to the model (**Fig.2B-E, red panels**). The prediction that $|a\text{-}c|$ would be larger in healthy subjects than in psychotic subjects was confirmed for lexical diversity (N), short-range recurrence (RE) and graph size (ASP), but not for long-range recurrence (LSC) (**Suppl. Table 7**). This occurred because LSC had lower *c* values in the psychotic sample than in the healthy sample, while *a* values were more similar across groups. Thus, the long-range recurrence deficit in psychotic subjects may reflect not a return to an immature pattern, but rather a developmental course that strays from the healthy profile from start.

In healthy subjects, word repetitions (RE) decreased exponentially within the first year of formal education, in parallel with a saturating increase in lexical diversity (N). Graph size (ASP) also increased, but with much slower dynamics that begins to saturate around the beginning of high school. Long-range recurrence (LSC) behaved similarly, with a characteristic time near the end of high school. To further test the null hypothesis of lack of temporal structure in the data, the temporal order of the samples was randomized 1,000 times and the graph attributes of this surrogate dataset were compared to real data. As shown in **Suppl. Table 8**, such disruption of temporal order abolished significant Spearman correlations (**Supplementary Fig. 1A**) and greatly reduced the $R^2$ of the exponential models (**Supplementary Fig. 1B**).

If memory reports from psychotic subjects are more disorganized than the reports of educated healthy adults, it is conceivable that their structure is also closer to that of random graphs [42]. To gain insight into the structural randomness of our samples, each graph was randomized 100 times by keeping the nodes and shuffling the edges (**Fig. 3A**). Normalizing each graph attribute by the corresponding mean random graph attribute, LSC and ASP from healthy controls with > 12 yE were significantly larger than in healthy controls with < 12 years of education (**Fig. 3B**). RE showed the opposite profile: Above random in healthy controls with < 12 yE, and near-random with >12 yE. None of these education-related differences in discourse structure were significant in psychotic subjects (**Fig. 3B**).

The results reveal different scales for the healthy maturation of distinct aspects of discourse structure, confirming the expectation of a protracted dynamics of characteristic times, which either precede or coincide with adolescence. That these changes span the entire period of regular schooling points to the importance of high school completion [43]. It also seems that education, not age, shapes the structural modification of discourse from early childhood to adolescence. This process requires time, but developmental time *per se* does not suffice without education. Overall, the results support the notion that the forces driving the organization of discourse are cultural, re-enforcing the expectation that a similar pattern should be observed in the historical record.

## Historical dynamics of discourse structure

Next we assessed whether the ontogenetic dynamics of graph attributes structurally resembles the historical development of the same attributes in texts from ~3,000 BC to 2,010 AC (**Fig. 4A**; left panel). For standardization, the analyses were performed in English. Mimicking the ontogenetic pattern, lexical diversity, graph size and long-range recurrence increased steadily over time across different traditions, while short-range recurrence decreased (**Fig. 4B-E; Suppl. Table 9**). Using 3,000 AC as the most parsimonious estimation of t=0 for the birth of written culture (**Suppl. Materials 4, 5A**), the historical textual data were remarkably well fit by the same model that described the ontogenetic data in healthy subjects (**Fig. 4B-E**). The null hypothesis of lack of temporal structure in the data was refuted by the same surrogation procedure described above (**Suppl. Table 10**, **Suppl. Fig. 2A,B**). As expected, *a-c* was positive for all graph attributes except RE, which was negative (**Suppl. Table 9**).

While the earliest texts show a near-random structure, later texts depart progressively from randomness. This is clear in a 2D plot of LSC and RE normalized by mean random values, which reconstitutes the temporal dynamics of the data based solely on structural properties (**Fig. 5A**). Indeed, 40% of the time variance among texts is explained by a single scalar combining normalized LSC and RE (**Fig. 5B**). A particularly interesting case is that of Hinduist literature, which evolved across 2,750 years from a primitive pattern of near-random long-range recurrence to its opposite (**Fig. 5C**; **Suppl. Material 5C,D).**

The exponentially saturating fits yielded characteristic times for the dynamics of graph attributes in literature (**Suppl. Table 9**). The results indicate that discourse structure began to mature much after the earliest written record, with a characteristic time of 1,503 BC (LSC), well within the middle Bronze Age. Another important period of maturation occurred at the end of the Bronze Age, with a characteristic time of 1,028 BC (RE) (**Suppl. Material 5D**). Interestingly, the saturation of lexical diversity and graph size is estimated to be in the distant future: 5,120 AC (N), and 98,873 AC (ASP).

Before the invention of writing, the ability to narrate real or fictional events was nearly exclusively mediated by oral storytelling. Short-range recurrence was likely favored because it facilitates rhyme and rhythm, as well as the memorization of short strings of words [44,45]. The need for attentive recall and the taste for reiteration is emphatically expressed in the words of the last king of the Sumerian city-state of Shuruppag in one of the earliest extant texts, possibly dating from before 3,000 BC: *"In those days, in those far remote days, in those nights, in those faraway nights, in those years, in those far remote years, at that time the wise one who knew how to speak in elaborate words lived in the Land; Shuruppag, the wise one, who knew how to speak with elaborate words lived in the Land. Shuruppag gave instructions to his son; Shuruppag, the son of UbaraTutu gave instructions to his son Ziudsura: My son, let me give you instructions: you should pay attention! Ziudsura, let me speak a word to you: you should pay attention!"* [46] (**Fig. 5A**).

However, a highly recursive structure hinders the communication of complex meaning, which requires long-range semantic context and imagetic schema [47], but is disrupted by short cycles [48]. Load restrictions on attention and working memory [49,50] must have limited the structural complexification of narratives for millennia. The invention of written text as an external support for memory allowed for a substantial increase in the size and complexity of the narratives, no longer constrained by the needs and strategies of memorization. This transformation seems to be well captured by our analysis. Ancient literature became structurally more complex as it developed, with an increase over time in the diversity of words employed, fewer repetitions of short-range word sequences and increasingly longer connected components. In particular, the dynamics of recurrence is characterized by a monotonic increase in range, likely reflecting the departure from oral to written discourse, the former strictly dependent on working memory, the latter much less so.

Computer science and mathematical modeling have been increasingly applied to archeological and historical research [40,51-56]. For text analysis across multiple live and dead languages and alphabets, this approach has the caveat of the need to use translations, mitigated here by the use of a single target language (English), and by the translation robustness of the differential diagnosis of psychosis based on graph analysis, which is nearly invariant across six major European languages including English [30]. To further investigate translation as a potential source of noise, untranslated original texts (N=28) were subjected to graph analysis for comparison with their English translations. Significant positive correlations were observed for N, RE and ASP (**Suppl. Fig. 4A**), but LSC showed no correlation due to a subset of Pre-Axial texts with substantially larger LSC in the English translations than in the originals (**Suppl. Fig. 4A**). As a consequence, the abrupt LSC increase at the Axial Age onset is even more marked in originals than in translations (**Suppl. Fig. 4B**). Overall, the dynamics of graph attributes in the original texts agrees with the results obtained for the larger sample of translated texts.

Unintended bias in the reference sample is another potential caveat: While our selection of classical texts is quite comprehensive, the sampling becomes increasingly arbitrary due to book popularization following Gutenberg's printing press ~1,440 AC. To address this criticism, 10 sets of 20 post-medieval texts were randomly sampled (**Suppl. Material 2**) and their graph attributes do not differ significantly from those of the reference sample (**Suppl. Fig. 3**). A further potential criticism is the particular choice of mathematical model. We chose to adjust the data to the simplest possible model, one that only presupposes linear dynamics that converges to a stable fixed point. This provides useful parameters to interpret the data, as indicated by the agreement with the dating of civilizational collapse between the Pre-Axial and Axial periods (**Suppl. Material 5B,D**).

## Structural randomness as a trace of immature discourse

Inferring the ancient mind based on a mathematical analysis of arcane records has an inevitable degree of speculation, but cognitive archeology gains depth when ancient literary data are compared to extant psychological data. The structural dynamics of historical texts shows similarity to the dynamics observed in healthy subjects, and most pre-Axial texts have graph attributes comparable to those measured in present-day reports from psychotic adults or healthy children. One way to interpret the data is to consider that ancient literature resembles psychotic speech. Another is to conclude that ancient written discourse is structurally comparable to verbal reports of present-day children. Both interpretations resonate with the notion that adult psychosis reflects primitive or infantile residues [57]. This is likely related to developmental limitations in working memory and attention [49,58], which subside with education [59,60]. Not surprisingly, limitations also observed in psychotic patients [61,62].

From a strictly structural point of view, cultural accumulation allowed for changes across 2.5 millennia that in healthy children take ~12 years of schooling. Surely Plato's writings were no adolescent material, being manifestly interested in adult topics. Yet, Plato's writings and other Axial classics are at par in structural complexity with verbal reports from modern-day healthy adolescents: Much closer to Voltaire than to Shuruppag (**Fig. 5A**). Childish or psychotic as it may, the written record abruptly reached a plateau of long-range recurrence around 800 BC, as shown by a moving window averaging of the data across all traditions (**Fig. 5D**). This sharp empirical transition, as well as the characteristic time for RE (1,028 BC), agrees very well with the cultural collapse between the end of the Bronze Age (~1,000 BC) and the onset of the Axial Age (~800 BC) (**Suppl. Material 5B,C,D**), when droughts, famine, plagues, war, invasions and natural cataclysms led to social disorganization, educational disruption, and literacy reduction [63]. Most interestingly, this transition represented a departure from near-random discourse structures (**Fig. 5D, right panel**).

The Axial Age has been challenged as a vague concept without empirical evidence [38,64]. However, a quantitative semantic analysis of Judeo-Christian and Greco-Roman texts revealed an increase in text similarity to the concept of "introspection" throughout the Axial Age [21]. Recently, statistical modeling attributed the timing of the Axial Age to economic development, not political complexity nor population size [39]. This has been interpreted as evidence that the intellectual blossoming of the Axial Age derived from changes in reward systems (from short to long term goals), rather than from changes in cognitive styles (from analogical to introspective)[39,40]. Our results argue for a complementary view: The economical prosperity of the Axial Age co-existed with a major change in discourse structure, with a contemporary parallel in the maturation of verbal reports that depends more on years of formal education than on biological age.

The characteristic times for the ontogenetic and historical development of graph attributes are summarized in **Figure 6**. Education-related cultural accumulation makes discourse less recursive and more connected at both the ontogenetic and historical levels, but the corresponding transformation paths are only partially overlapping. While the monotonic dynamics in both datasets are overall quite similar (compare **Figs. 1C, 2** and **4**), the temporal order of saturation for specific graph attributes differs across datasets.

Ontogenetically, graph attributes related to short-range recurrence and lexical diversity begin to stabilize in the first school year, as expressed in a wider use of an expanding vocabulary and less use of mnemonic resources to organize speech. Then, mostly during high school but with large inter-individual variation, graph size and long-range recurrence saturate, and graph attributes evolve towards the typical adult profile. The data point to a hierarchical development of discourse structure, by which we depart from an initial pattern of fragmented word segments dominated by short-range connections to a learned pattern of globally connected word strings.

Historically, the earliest maturation of discourse structure occurred for the increase in long-range recurrence during the middle Bronze Age, followed by the decrease in short-range recurrence near the onset of the Axial Age. Similarly to the ontogenetic data, a decrease in short-range recurrence is an early marker of maturation in literature. However, lexical diversity and graph size follow a distinct path, not stabilizing until much beyond the present. These differences are likely related to the fact that the historical data was not produced by children, but by educated adults of the cultural elites of yore. Still, the different paths reach similar outcomes. The results imply that, at any given time, it is the educated subject able to create literature – the writer – who will push the envelope of discourse structure, and therefore school curriculum. The fine-grained developmental trajectories for graph attributes are different for ontogenesis and history because of the many intrinsic differences between these processes, including the fact that they correspond in the latter to the maximum found in the population, while in the former they simply measure the degree of adherence to the current educational canon.

First established in ancient Sumer [65,66], schools foster the education of those who will instruct younger generations through written language. Literacy acquisition is associated with important anatomical and physiological changes in neocortical organization, including robust lateralization [6,67-70]. Given the association between psychosis and lack of lateralization [71-74], the present results suggest that the lateralization associated with literacy may have shaped the mental processes underlying the development of literature. Significantly, psychotic symptoms were not considered pathological until around 360 BC, when Plato defined insanity as a serious disease (Theaetetus 158), "totally out of place in a well-regulated State" (Laws 461) [75]. While the complex discourse structure of healthy adults owes more to nurture than to nature, education does not do its work in psychotic subjects. When cognitive development is impaired by disease, nature trumps nurture. Despite exposure to education, psychotic subjects retain a linguistic structure akin to that of infantile speech, failing to mature in complexity and remaining closer to a near-random structure. The historical parallel of a psychotic breakdown with cognitive decline is given by the cultural collapse at the end of the Bronze Age, which coincides with the resurgence of literature with increased short-range recurrence and decreased long-range recurrence. The structural randomness of long-range connections seems therefore to represent an immature trace of the human mind, at the level of the individual as well as historically.

## Figure Legends

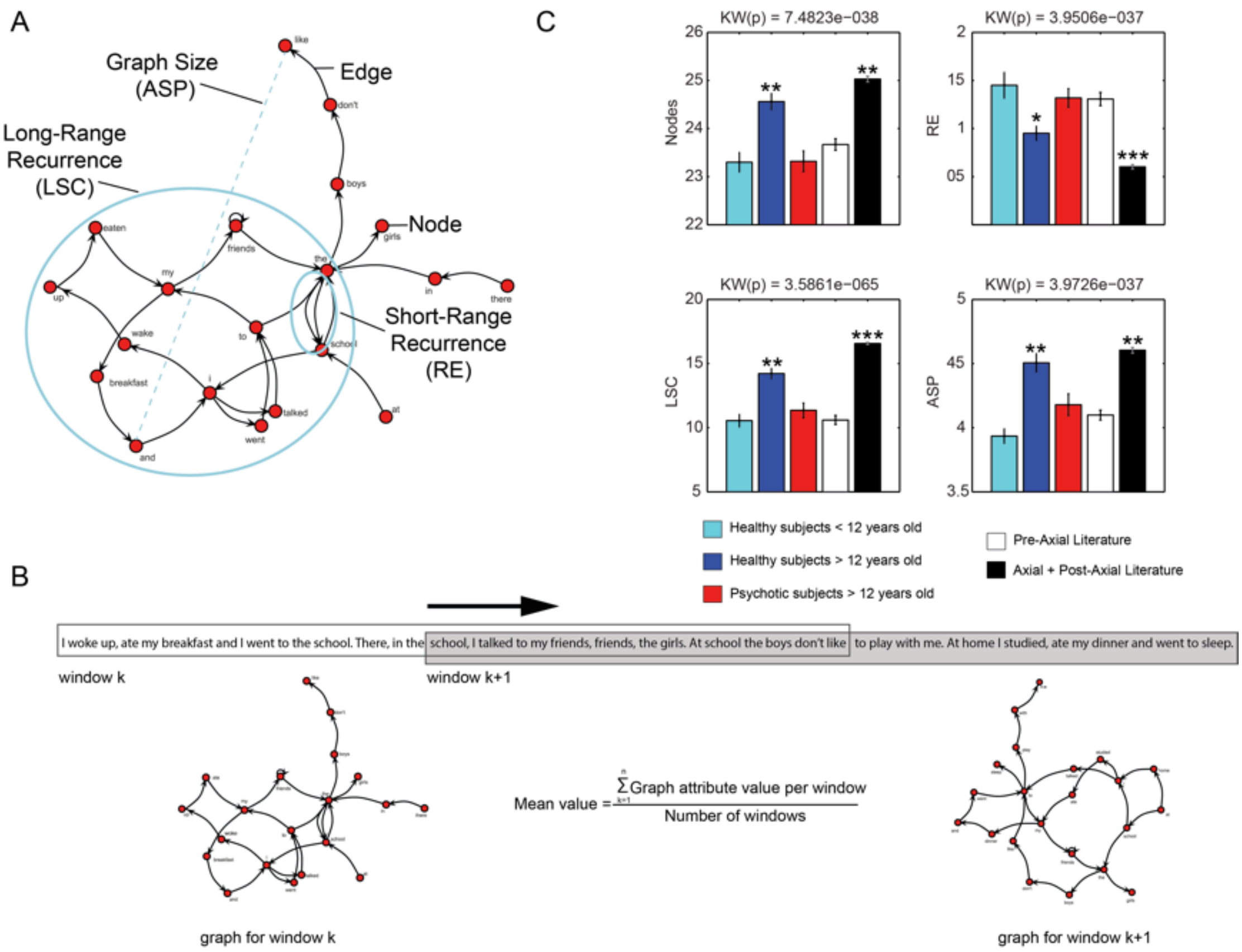

**Figure 1: Verbal reports from healthy children and psychotic adults are structurally similar to Pre-Axial literature, while reports from healthy adults resemble literature from Axial and Post Axial periods. A)** The graph attributes investigated comprised lexical diversity (N), long-range recurrence (LSC), short-range recurrence (RE) and graph size (ASP) [29,30]. Red circles indicate nodes, black arrows indicate edges. **B)** Moving windows (length = 30 words, 50% overlap) were used to calculate mean values per graph for the different attributes investigated. Graphs of different sizes can be quantitatively compared using this procedure. **C)** Graph attributes from psychotic subjects are not significantly different from those of healthy children and Pre-Axial literature (**Suppl. Table 3**). Mean ± SEM, * for statistically significant differences ($p<0.05$ corrected for multiple comparisons) with Pre-Axial Literature, ** for the same differences plus healthy children < 12 years and psychotic subjects, *** for the same differences plus healthy subjects > 12 years.

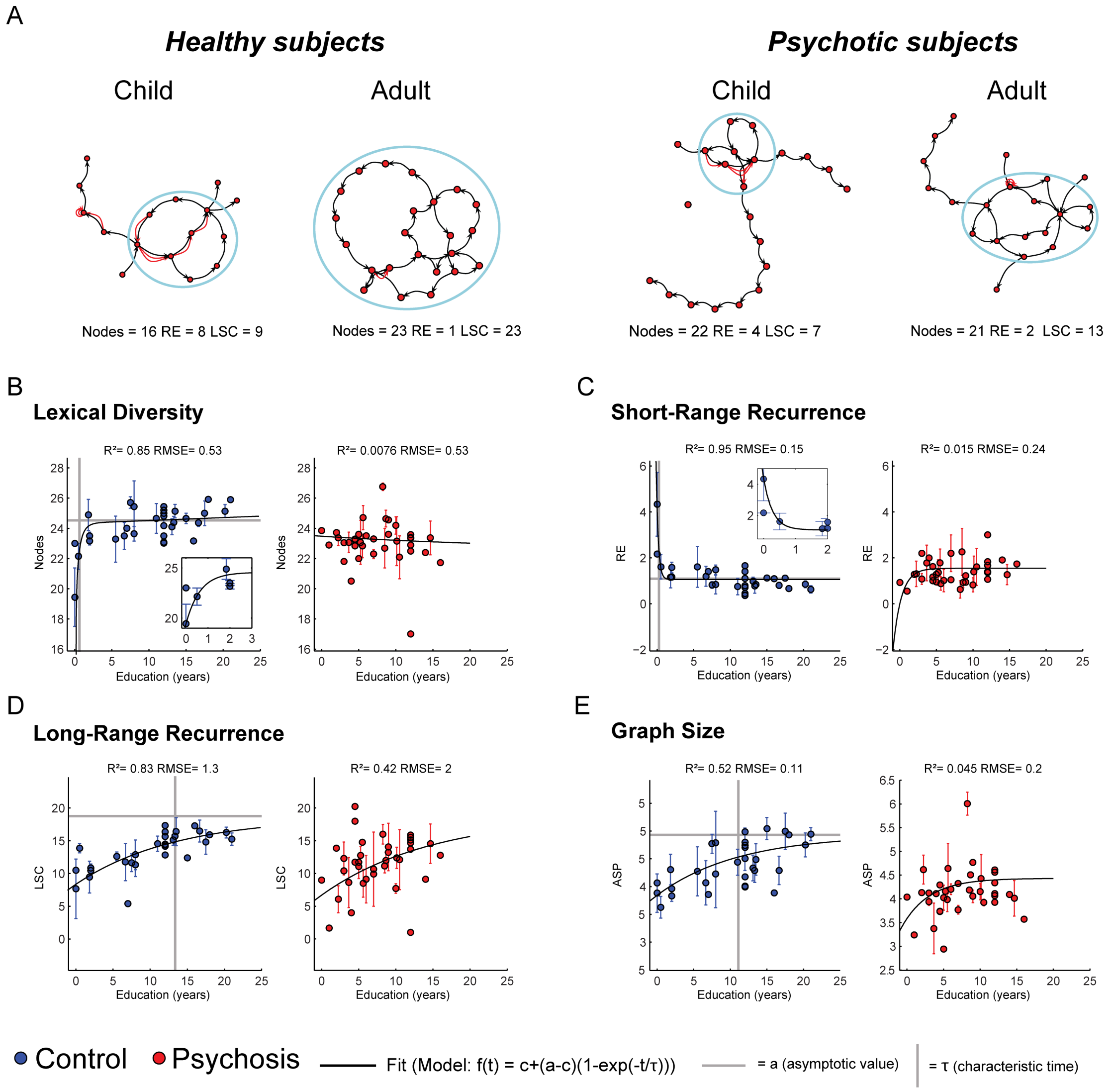

**Figure 2: The structure of memory reports matures with years of education in healthy subjects, but not in psychotic patients. A)** Representative examples of graphs from healthy and psychotic subjects, as children or adults. Light blue perimeters indicate LSC. **B)** Lexical diversity as a function of years of education (yE) for control (left) and psychotic (right) subjects. Similar plots for **C)** Short-range recurrence, **D)** Long-range recurrence, and **E)** Graph size. For significant Spearman correlations, characteristic years of education and asymptotic value are indicated by vertical and horizontal dashed lines, respectively. RMSE indicates root-mean-square error.

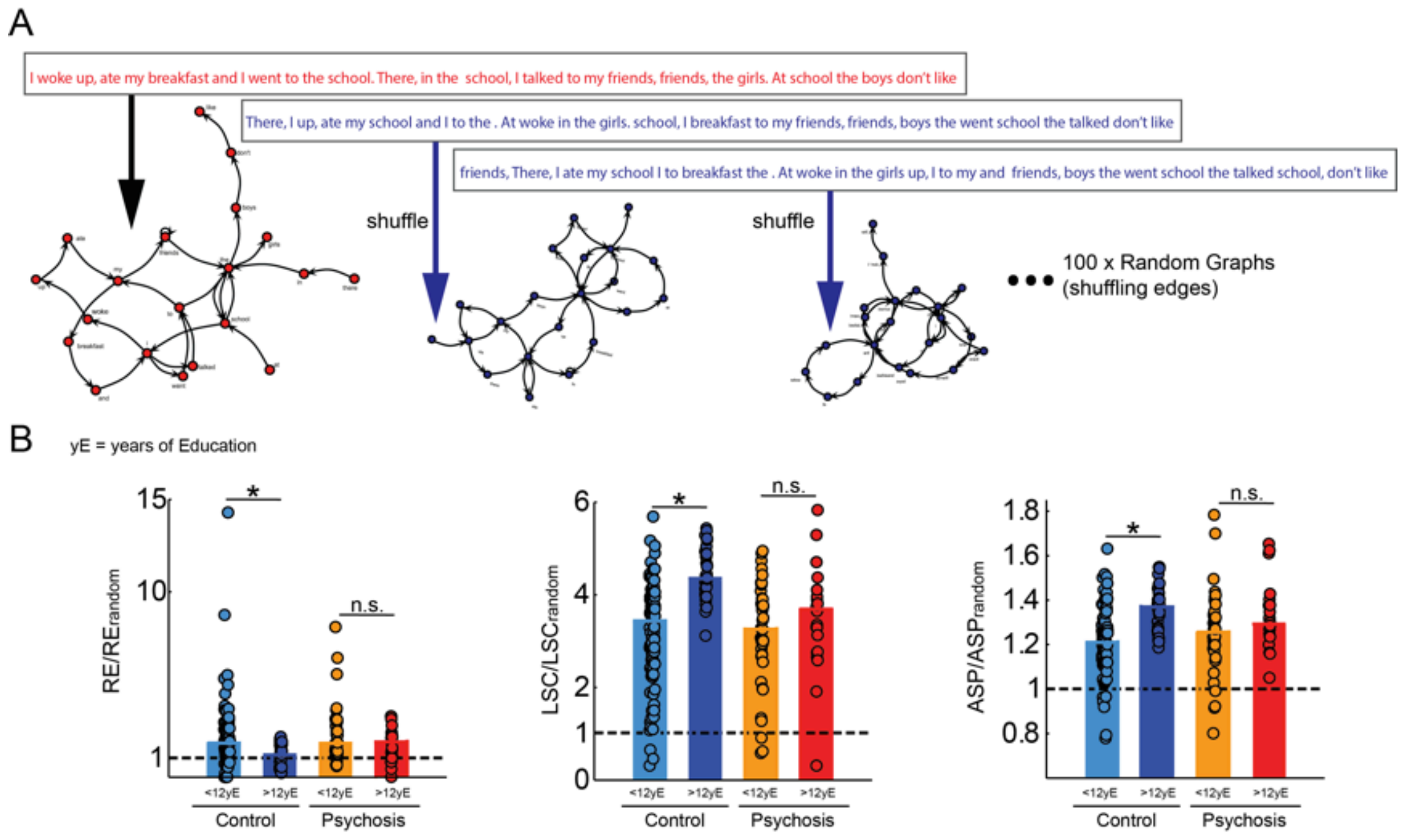

**Figure 3: Memory reports from psychotic subjects have a near-random structure. A)** To calculate mean random values, each graph was randomized 100 times by shuffling edges and preserving nodes. Graph attributes were calculated for each random graph and averaged to compose the denominator of the ratio shown as normalized graph attribute in the next panel. **B)** The graph attributes of each individual report were normalized by the corresponding mean random value, and the data were sorted according to more or less than 12 yE. Control subjects showed significant differences between subjects below or above 12yE, but psychotic subjects did not; * for $p<0.05$ corrected for multiple comparisons, n.s. for non-significant differences (Wilcoxon test, $\alpha=0.0028$). Control subjects > 12 yE also showed significant differences from psychotic subjects < 12 yE for all graphs attributes, and from psychotic subjects > 12 yE for LSC.

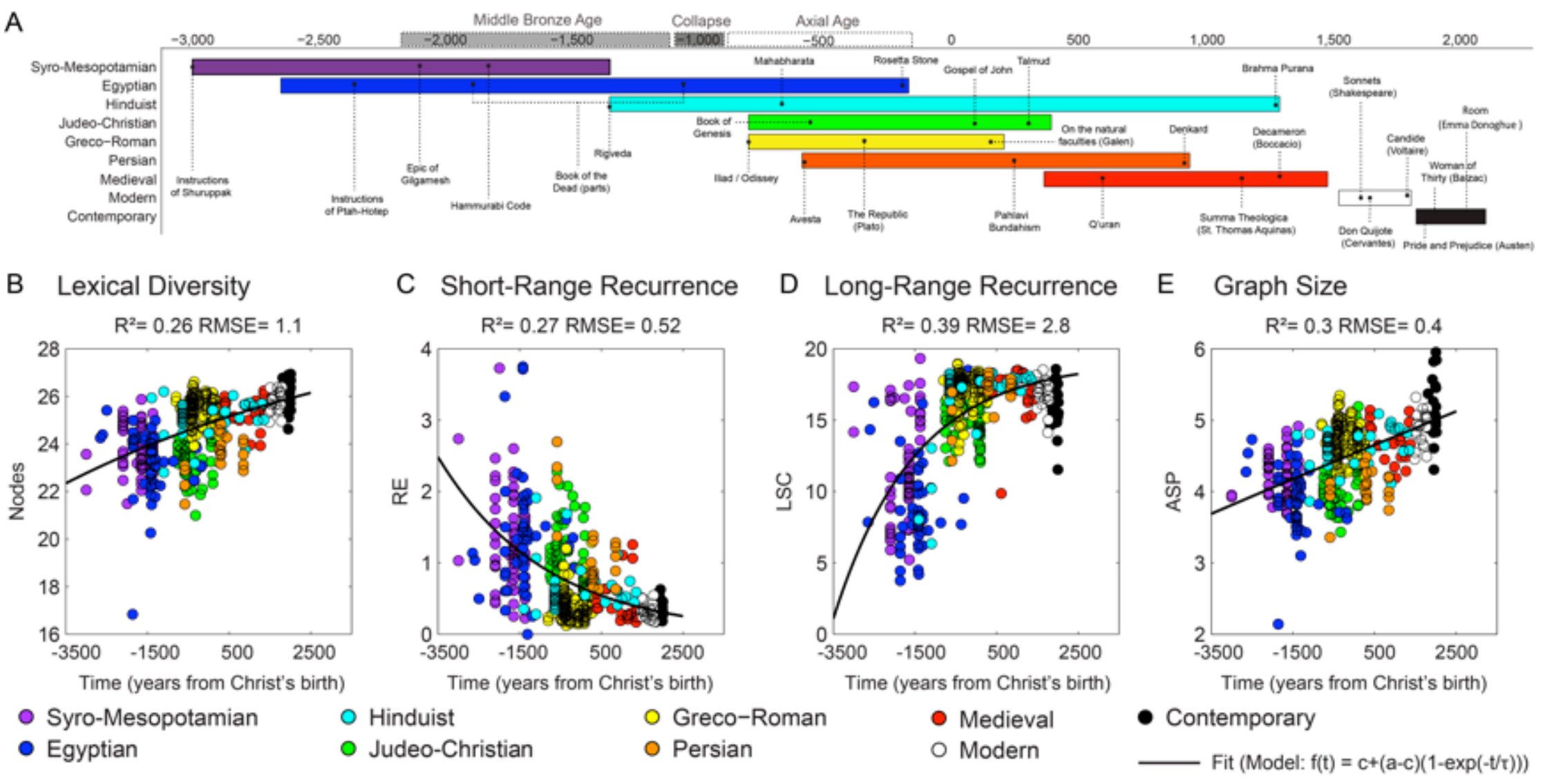

**Figure 4: The historical development of literary structure mimics the ontogenetic dynamics. A)** A corpus of 448 representative texts across 9 Afro-Eurasian literary traditions spanning 5,500 years was investigated by graph analysis as in Figure 1. **B)** Lexical diversity increased monotonically over time, while **C)** Short-range recurrence showed the opposite dynamics. **D)** Long-range recurrence and **E)** Graph size increased over time. The data are well explained by the exponentially saturating model.

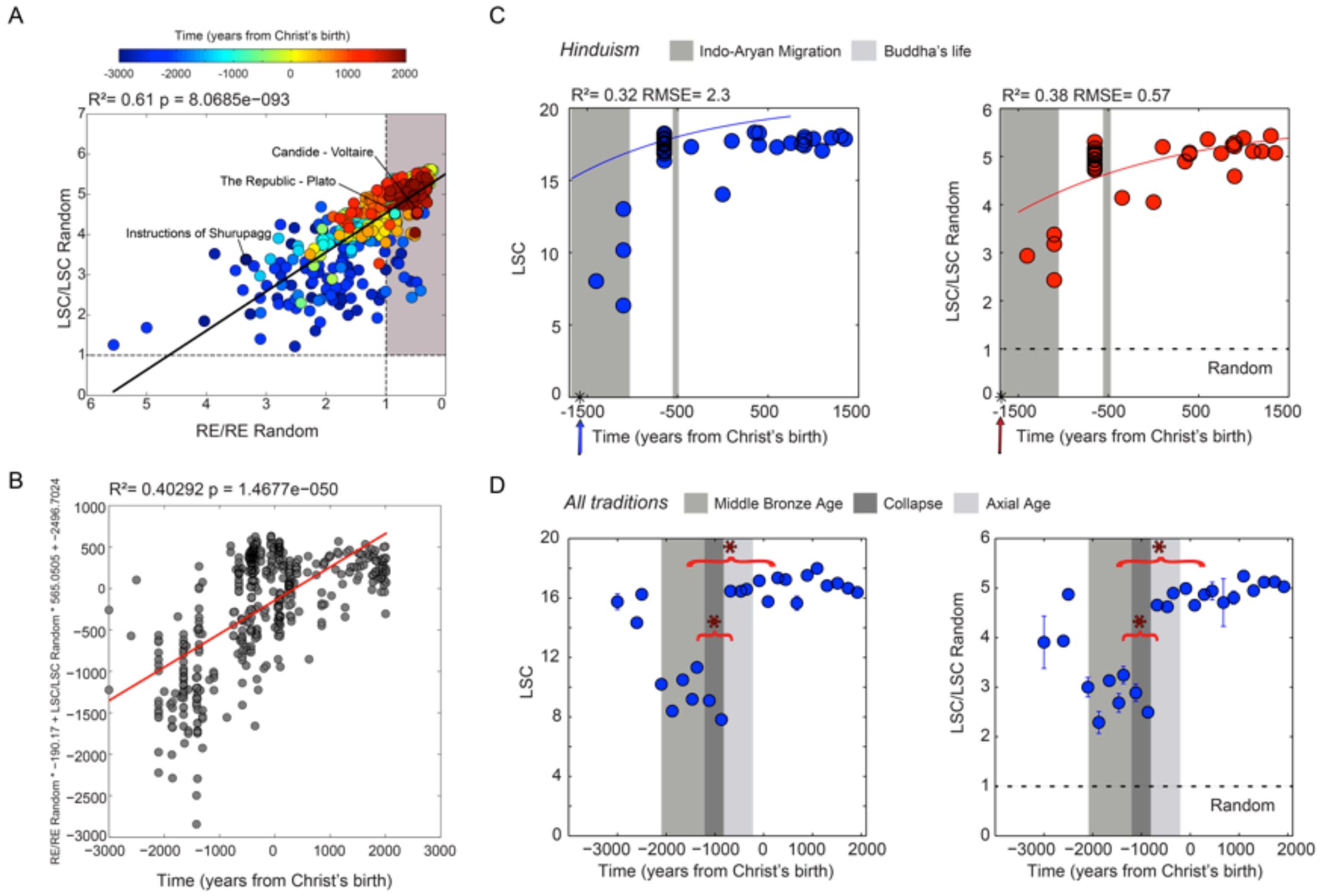

**Figure 5: The maturation of literary structure reflects historical time. A)** LSC and RE normalized by mean random values provide a smooth distribution of texts along the "arrow of time". Normalization consisted of dividing each data-point by its corresponding mean random value. **B)** A linear combination of LSC and RE normalized by mean random values shows strong significant correlation with historical time. **C)** Long-range recurrence measures exponentially saturate over time in the Hinduist literature, with characteristic times within the period estimated for the Indo-Aryan migration (**Suppl. Material 5B,C,D**). **D)** The structural long-range recurrence of literary texts shows a marked transient, with a sharp decrease after the Bronze Age collapse (~1,000 BC), followed by a saturating increase at the onset of the Axial Age (800 BC). Left panel shows LSC, right panel shows LSC/random plotted as moving window averages across all traditions, with windows of 200 years without overlap (mean ± SEM). * for $p<0.05$ corrected for multiple comparisons. There is striking agreement between this empirical transition and the dating of civilizational collapse and revival between Pre-Axial and Axial periods (**Suppl. Material 5B,D**).

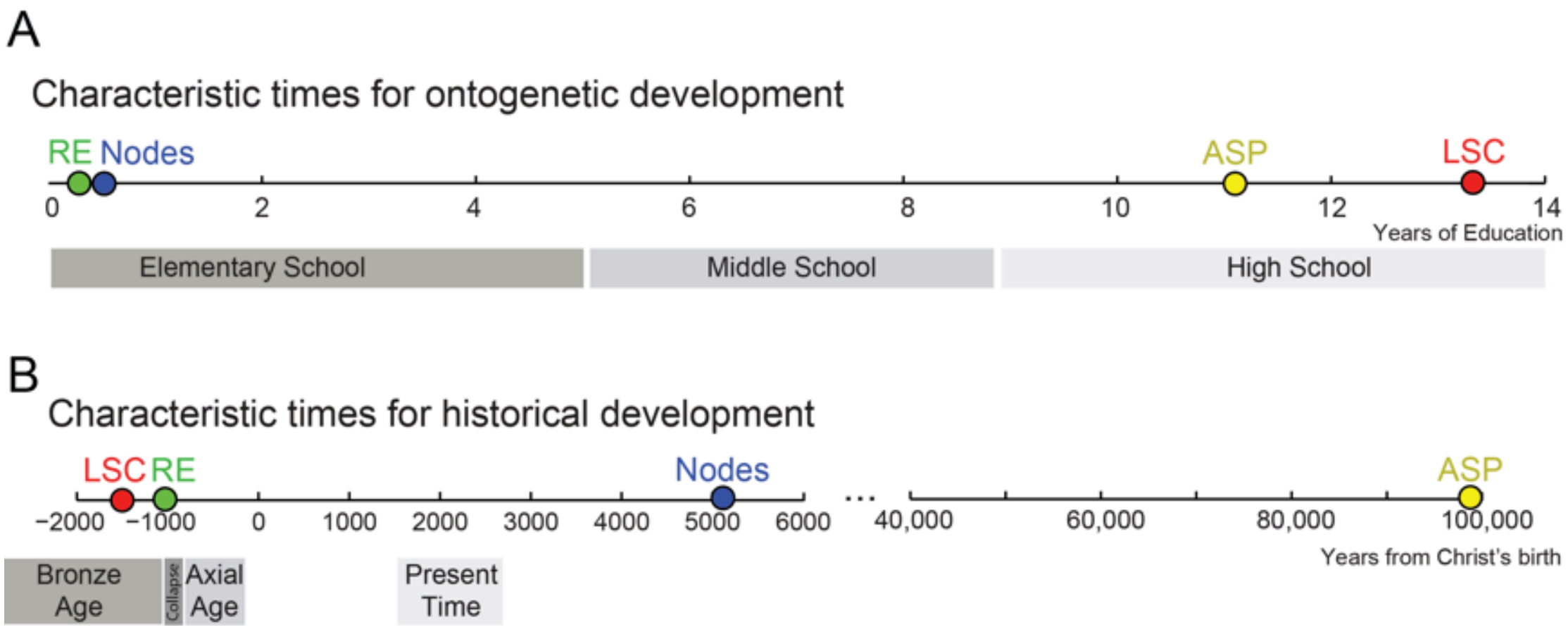

**Figure 6: Characteristic times for the ontogenetic or historical maturation of graph attributes over time.** Colors indicate graph attributes related to lexical diversity (blue), short-range recurrence (green), long-range recurrence (red), and graph size (yellow). The temporal order of maturation for specific graph attributes differs between ontogenetic and historical data. In the latter, note that Nodes and ASP are predicted to continue maturing further in the future.

## Suppl. Fig. Legends

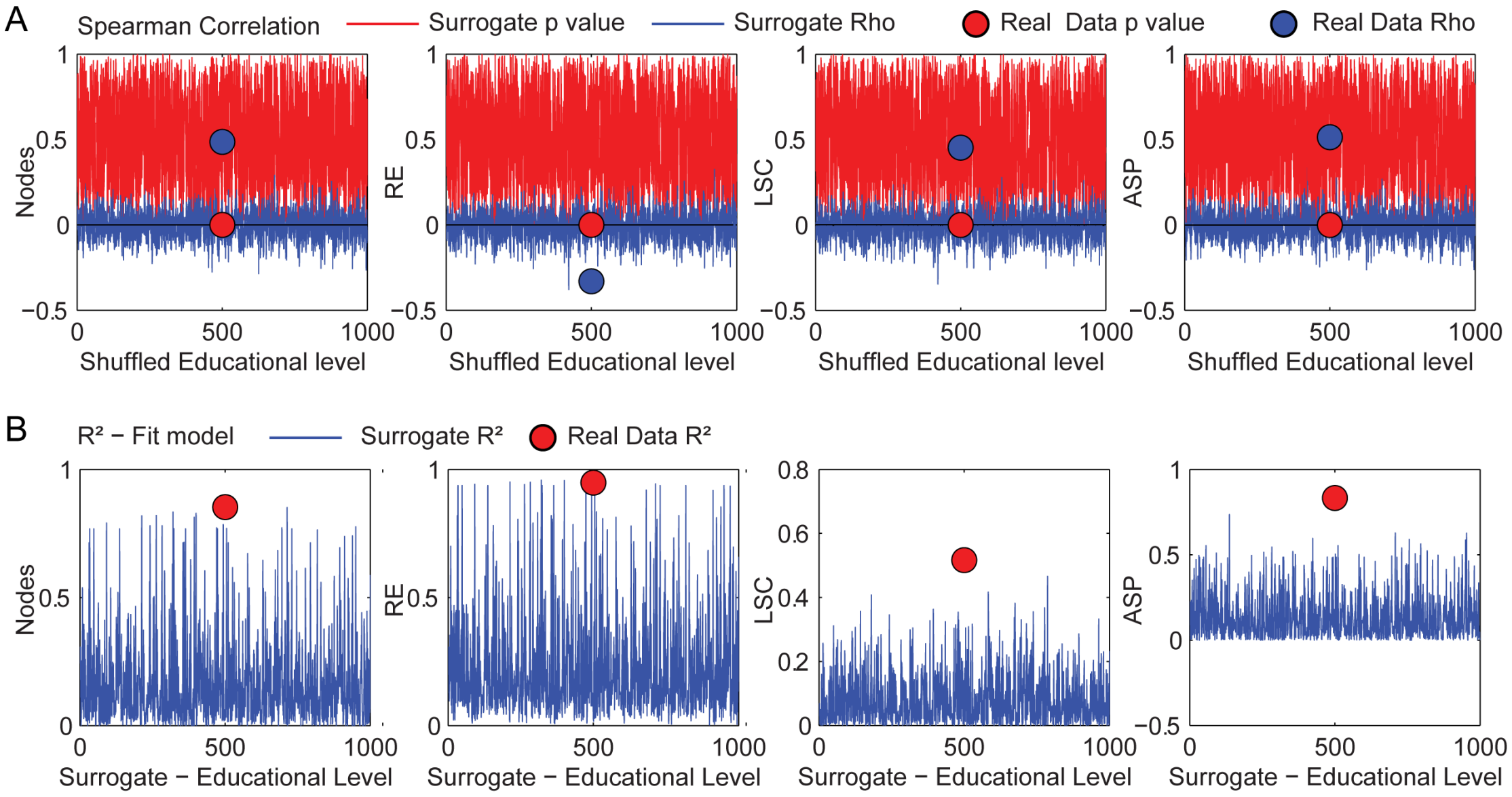

**Suppl. Fig. 1: Ontogenetic data randomized for educational level (1,000 surrogations) do not correlate with graph attributes.** A) Spearman correlation performed with 1,000 shuffled educational level with graph attributes indicated with lines (surrogate Rho indicated as blue line and surrogate p values indicated as red line), compared with real data correlation, performed with real educational level, indicated by dots (real data Rho indicated by blue dot and real data p value indicated by red dot. B) Asymptotic fit performed with 1,000 shuffled educational level with graph attributes indicated with lines (surrogate $R^2$ indicated by blue line), compared with real data fit, performed with real educational level, indicated by dot (real data $R^2$ indicated by red dot).

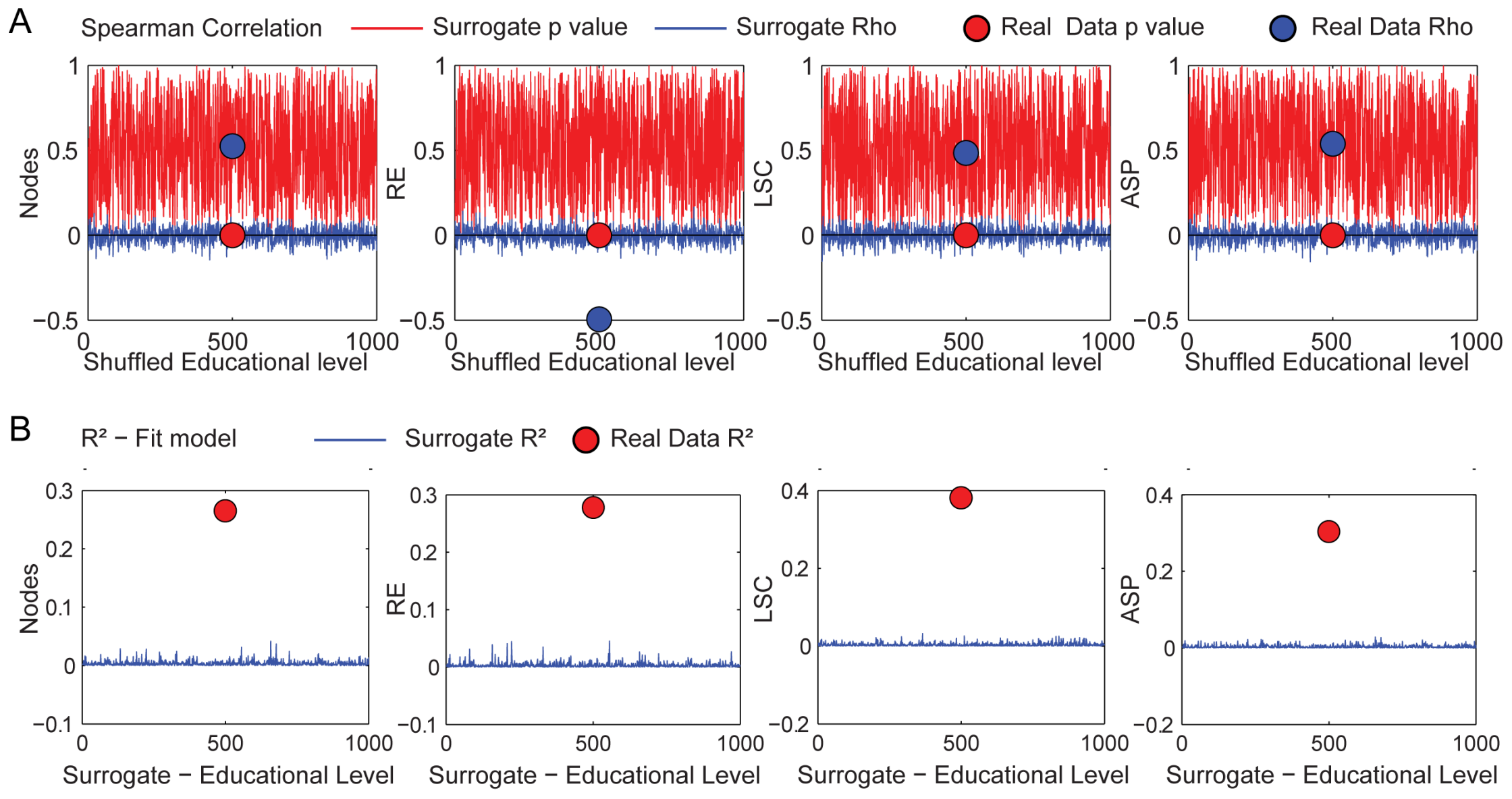

**Suppl. Fig. 2: Historical data randomized for time (1,000 surrogations) do not correlate with graph attributes.** A) Spearman correlation performed with 1,000 shuffled historical data with graph attributes indicated with lines (surrogate Rho indicated as blue line and surrogate p values indicated as red line), compared with real data correlation, performed with real historical data indicated by dots (real data Rho indicated by blue dot and real data p value indicated by red dot. B) Asymptotic fit performed with 1,000 shuffled historical data with graph attributes indicated with lines (surrogate $R^2$ indicated by blue line), compared with real data fit, performed with real historical data, indicated by dot (real data $R^2$ indicated by red dot).

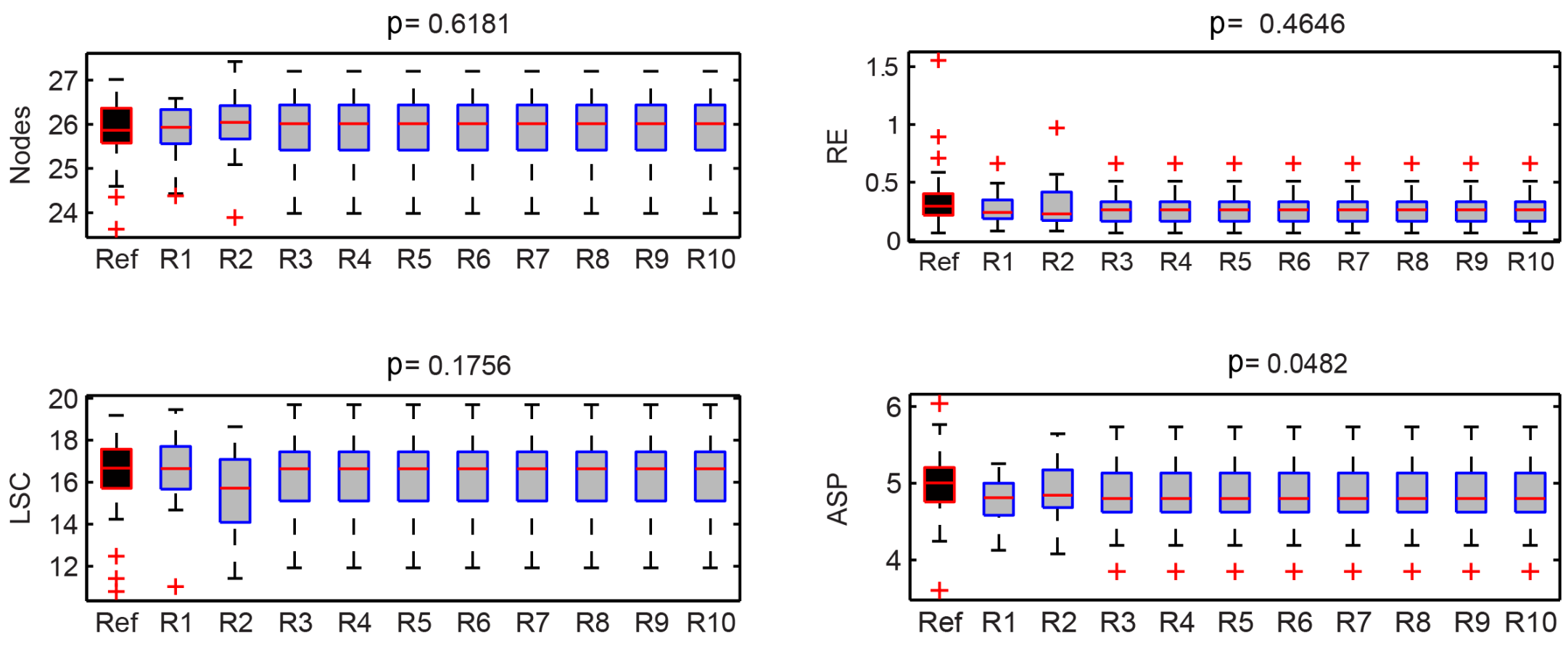

**Suppl. Fig. 3: Graph attributes of the reference sample of post-medieval texts do not differ from those of random samples.** Compare results from the reference sample (Ref; black boxplots) and 10 samples of 20 post-medieval texts randomly chosen from the Gutenberg Project digital library (R1-R10, gray boxplots). P values for Kruskal-Wallis tests corrected for 4 comparisons ($\alpha$=0.0125).

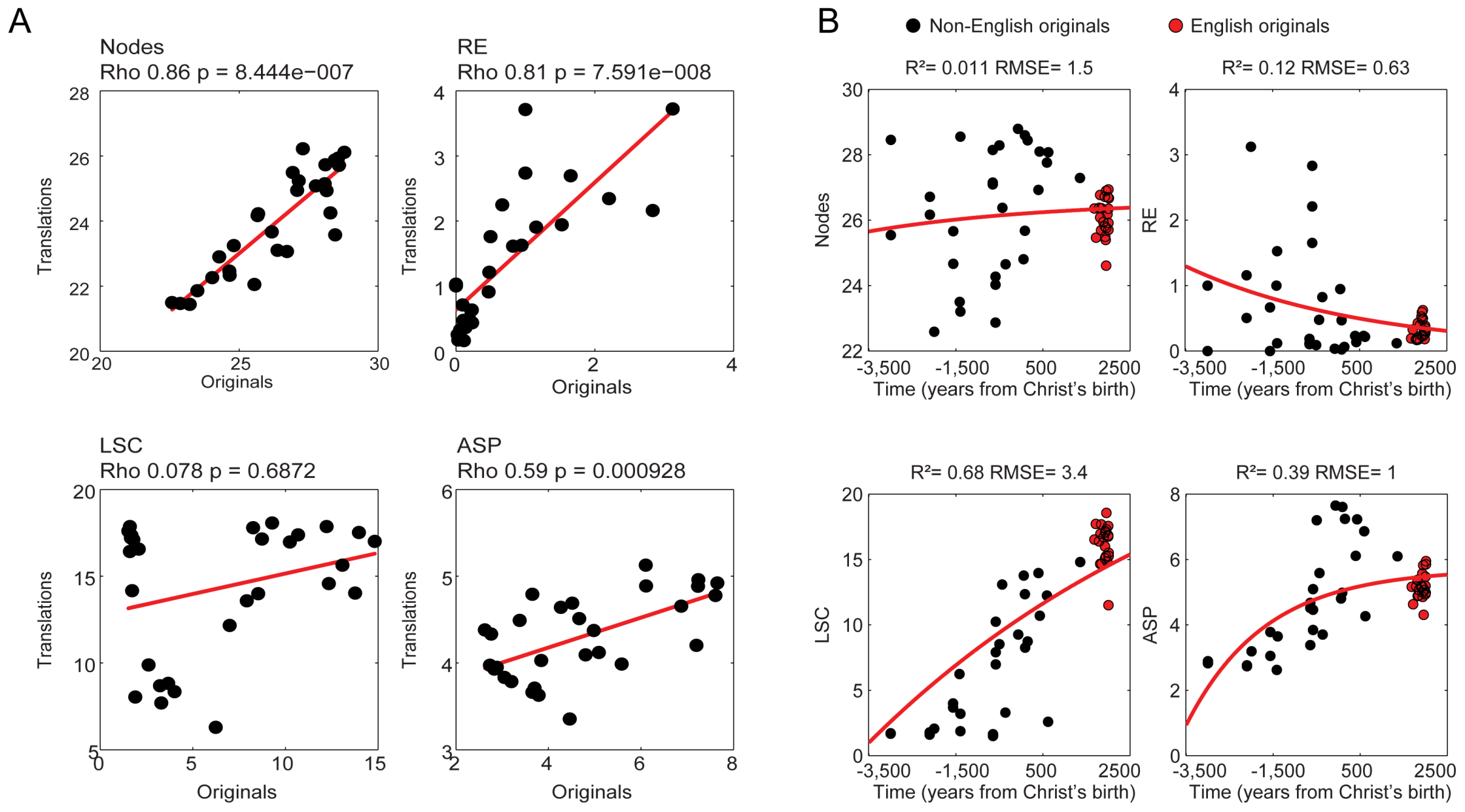

**Suppl. Fig. 4: Comparison of graph attributes between original and translated texts. A)** Nodes, RE and ASP were significantly correlated between originals and translations. LSC was not, due to a subset of Pre-Axial texts in the top left corner of the plot, with much larger LSC in the translations than in the originals. **B)** The dynamics of graph attributes in original texts shows monotonic changes quite similar to those observed in translated texts (compare with **Fig. 4**). Note the structural clustering of English originals.

## ***<u>Supplementary Information Index:</u>***

**Supplementary Material 1: Ontogenetic Data.**

The sample (data pooled from [1-4] plus new samples) comprised clinical oral interviews from 200 individuals (135 without any diagnosis of psychiatric disorder, and 65 independently diagnosed by the standard DSM IV ratings SCID [5] with psychotic symptoms as schizophrenic (S) (N=36) or bipolar type I (B) (N=29) (**Suppl. Table S1**). Also applied were two standard psychometric scales, the ''Positive and Negative Syndrome Scale'' (PANSS) [6] and the ''Brief Psychiatric Rating Scale'' (BPRS) [7], and a socioeconomic-clinical questionnaire (with information regarding age, sex, family income, educational level, marital status, disease duration and onset). This study used data from two protocols approved by the Research Ethics Committee of the Federal University of Rio Grande do Norte (permits #102/06-98244 and #742.116). Signed informed consent was obtained from each subject and also from a legal guardian when necessary. The exclusion criteria were any neurological condition or alcohol/drug abuse. The analysis of memory reports focused on answers to three open questions, namely requests for reports on one recent dream, on waking activities in the previous day, and about a negative affective image shown for 15 seconds immediately before the request. The negative image was selected from a widely validated affective images database [8] [4]. For each subject, the three reports were concatenated and the final text was represented as a word graph (**Figure 1A**).

**Supplementary Material 2: Literary Data.**

Bibliography Selection and Edition: 448 representative prose texts translated to English or written in English were extracted from the public domain of internet or kindly provided by their authors were converted to .txt extension and edited to remove prefaces, notes, comments, line breaks, page/tablet numbering and publisher information. Paragraphs were preserved. All text editing procedures performed with Matlab and Notepad++ software.

Control for arbitrary selection of post-medieval texts: To compare with our literary sample, additional 200 texts comprising 10 random sets of 20 post-medieval texts were selected using the search engine "Random Page" on the digital library Project Gutenberg (https://www.gutenberg.org/ebooks/search/?sort_order=random). Plays, poetry and non-English versions were excluded. For this control, only the initial 1,000 words of each text were analyzed.

Untranslated originals: As a control for translation effects, 50 untranslated texts were also analyzed (28 non-English texts and 22 English originals already included in the initial sample).

Text Dates: Text dating information was obtained by exact dating (1), average dating within a reference period (2), average dating within the period of the author's life (3).

A total of 676 texts were analyzed. Text identification, time intervals, and dating are detailed in **Suppl. Table 2**. Text sources included the Digital Egypt of the University College London (http://www.ucl.ac.uk/museums-static/digitalegypt/), the Electronic Text Corpus of Sumerian Literature of the University of Oxford (http://etcsl.orinst.ox.ac.uk/), Project Gutenberg (www.gutenberg.org), The Internet Classics Archive of the Massachusetts Institute of Technology (http://classics.mit.edu/), the Avesta Zoroastrian Archives (http://www.avesta.org/), and the Internet Sacred Text Archive (http://www.sacred-texts.com/).

**Supplementary Material 3: Graph Analysis.**

Graph analysis was performed using the free software *SpeechGraphs* (http://www.neuro.ufrn.br/softwares/speechgraphs). For memory reports as well as literary texts, average graph attributes were calculated across each graph using moving windows of 30 words with 50% of overlap [2], i.e. steps of 15 words (**Figure 1B**). A total of 4 average graph attributes were calculated for each text file, comprising lexical diversity (Nodes=N), short-range recurrence (RE = repeated edges= RE), long-range recurrence (largest strongly connected component = LSC) and graph size (ASP = average shortest path). To estimate randomness levels, each 30-word window was shuffled 100 times so as to keep the same words but change their order (**Figure 3A**). This procedure is equivalent to a random permutation of edges [9]. Graph attributes of randomized word windows were then averaged and used to normalize the original average data (**Figures 3B, Fig. 5**). To cope with computational cost, texts above 50,000 words were trimmed to this maximum. Data analyzed in Excel and Matlab software.

**Supplementary Material 4: Exponential model**

In order to study the dynamics of graph attributes across different educational levels or across time in literature, the following model was used:

$$f(t) = c+(a-c)(1-\exp(-t/T))$$

where

$a$ is the maximum asymptotic graph attribute value
$c$ is the initial graph attribute value
t is time
T is characteristic time to reach saturation.

The function is the solution to a linear differential equation of first order:

$$df/dt = (1/T)(a-f) \text{ with initial condition } f(t=0)=c.$$

For memory reports we used as input data the average graph attribute from all individuals with the same age, and weighted the model for the standard error of the mean. For literary data we used a non-weighted model. To better adjust the fit, we considered lowers and uppers points to each coefficient, according to the maximum and minimum value expected for each graph attribute and for time (years of education or historical time).

| Coefficient | Lower point / rationale | Upper point / rationale | Start-point |
|---|---|---|---|
| a | 0 / no graph attribute can be smaller than 0 | 30 for N, and LSC (graph attribute counted by number of nodes) / maximum number of nodes for 30-word graphs) | Maximum observed value |
| | | 29 for RE and ASP (graph attributes counted by number of edges) / maximum number of edges for 30 word graphs) | |
| T | 0 for education / illiterates | 30 for education / post-doctoral level | 12 years of education / High school level |
| | -3000 for historical time / beginning of writing | Infinite for historical time / future | -800 / Axial Age |
| c | 0 / no graph attribute can be smaller than 0 | 30 for N, and LSC (graph attribute counted by number of nodes) / maximum number of nodes for 30-word graphs) | Minimum observed value |
| | | 29 for RE and ASP (graph attributes counted by number of edges) / maximum number of edges for 30 word graphs) | |

**Supplementary Material 5: Historical events of interest**

**A)** **The birth of literature** occurred in Afro-Eurasia during the early Bronze Age, in the context of the first major civilization merge, involving Indo-European and Semitic populations. Proto-Indo-European originated in west-central Asia 9,500 to 6,000 years ago, spawning since then to Europe and most of Afro-Asia as the multiple Indo-European languages [10-12] co-evolved with branches of the Afro-Asiatic linguistic family [13]. Cultural and linguistic diversity are estimated to have peaked during the Neolithic and declined afterwards [14,15]. Around 3,000 BC writing created the capacity for reliable communication across space and time, as the historical record began [16]. Population growth, migrations and military conquests began to periodically unify larger and larger groups of people around similar cultural kernels [17-20].

**B)** **The Axial Age (800-200 BC)** was marked by civilization blossoming in multiple Eurasian sites, including Athens, Rome, Babylon, and the Persian, Macedonian and Mauryan Empires [21-29]. Many fundamental texts of ancient literature date from this period (e.g. Iliad, Odyssey, The Republic, Book of Genesis, Avesta, Mahabharata). Multicultural development and integration was accelerated by the consolidation of alphabetic writing, new literary traditions and the foundation of the first high-level educational institutions, such as Plato's Academy and the Library of Alexandria in the 4th century BC. By 326 BC, when Alexander invaded northern India, Indo-European and Afro-Asiatic languages were developing sympatrically, with shared aspects of literature, religion, govern, trade and money [30,31].

**C)** Civilizations fell and rose in rapid succession at the end of the **early Bronze Age**, marked by severe aridification [32]. For instance, the collapses of the Old Egyptian Kingdom (~2,181 BC), and of the Akkadian Empire in Mesopotamia (~2,154 BC) were soon followed by empire reunification in Egypt (~2,055 BC) and Mesopotamia (~2,025 BC for Assyria and ~1,760 BC for Babylon) [33-35]. On the East, major urban centers dating from before 3,000 BC such as Mohenjodaro and Harappa, began to collapse by ~1,900 BC. The decay of the Indus valley civilization was followed by an early migration of Indo-Aryan populations into northwestern India between 1,800 BC and 1,500 BC [36,37]. Together with several other examples, these events mark the end of early Bronze Age and the onset of middle Bronze Age in Afro-Eurasia [38,39].

**D) The end of Bronze Age** is marked by a long list of city-states that collapsed or began to fade in the West at the dawn of the first millennium BC [40,41], including Knossos (~1,100 BC), Homeric Troy (Herodotus ~1,250 BC, archaeological Troy VII: ~950 BC), Mycenae (~1,200 BC), Ugarit (~1,190 BC), Megiddo ~1,150 BC, and Babylon (~1,026 BC). Collapses also occurred in the empires of Egypt (~1,100 BC) and Assyria (~1,055 BC). By 1,200 BC Indo-Aryan groups were penetrating eastward into the Ganges plains, and by ~1,000 BC the transition from semi-nomadic pastoral to settled agricultural Vedic societies was consolidated [36,38,42-45].

**Supplementary Table 1: Demographic characteristics of healthy and psychotic subjects.**

| | | Psychotic Subjects | | Controls |
|---|---|---|---|---|
| **Demographic Characteristics** | | **Schizophrenia** | **Bipolar** | |
| **N** | | 36 | 29 | 135 |
| **Age** | | 28.25 ± 6.75 (9 - 57 yo) | 31.07 ± 8.24 (7 - 58 yo) | 14.92 ± 6.22 (2 - 56 yo) |
| **Sex** | **Male** | 83% | 59% | 49% |
| | **Female** | 17% | 41% | 51% |
| **Years of Education** | | 11.96 ± 4.35 (0 - 16 y) | 15.20 ± 4.94 (0 - 16y) | 11.65 ± 6.40 (0 - 23y) |
| **Psychiatric Assessment** | | **Schizophrenia** | **Bipolar** | |
| **Disease Duration** | | 12.31 ± 12.62 | 8.28 ± 9.81 | ... |
| **Psychometric Scales** | **BPRS** | 16.81 ± 6.42 | 15.28 ± 7.18 | ... |
| | **PANSS** | 69.69 ± 14.79 | 62.45 ± 15.74 | ... |

## Supplementary Table 2: Identification and dating of literary texts.

| Highlighted texts analyzed in both original and translated versions | Dating method: 1 -Known date; 2 - middle of historical period when the text was written; 3 - middle of author's lifespan | | | |
|---|---|---|---|---|
| **Text identification** | **Tradition** | **Original Language** | **Dating method** | **Estimated date** |
| **Hymn to Enlil** | Syro-Mesopotamian | Sumerian | 1 | -3000 |
| **Instructions of Shuruppak** | Syro-Mesopotamian | Sumerian | 1 | -3000 |
| The instruction of kegemini | Egyptian | Middle Egyptian Hieroglyphs; Arcaic Hieratic | 2 | -2601 |
| The instruction of Ptah hoted | Egyptian | Old Egyptian hieroglyphs | 2 | -2501 .5 |
| Enmerkar and En suhgir ana | Syro-Mesopotamian | Sumerian | 1 | -2100 |
| **Enmerkar and the lord of Aratta** | Syro-Mesopotamian | Sumerian | 1 | -2100 |
| Gilgamesh 1 | Syro-Mesopotamian | Sumerian | 1 | -2100 |
| Gilgamesh 2 | Syro-Mesopotamian | Sumerian | 1 | -2100 |
| Gilgamesh 3 | Syro-Mesopotamian | Sumerian | 1 | -2100 |
| Gilgamesh 4 | Syro-Mesopotamian | Sumerian | 1 | -2100 |
| Gilgamesh 5 | Syro-Mesopotamian | Sumerian | 1 | -2100 |
| Gilgamesh 6 | Syro-Mesopotamian | Sumerian | 1 | -2100 |
| Gilgamesh 7 | Syro-Mesopotamian | Sumerian | 1 | -2100 |
| Gilgamesh 8 | Syro-Mesopotamian | Sumerian | 1 | -2100 |
| Gilgamesh 9 | Syro-Mesopotamian | Sumerian | 1 | -2100 |
| Gilgamesh 10 | Syro-Mesopotamian | Sumerian | 1 | -2100 |
| Gilgamesh 11 | Syro-Mesopotamian | Sumerian | 1 | -2100 |
| Gilgamesh 12 | Syro-Mesopotamian | Sumerian | 1 | -2100 |
| **Lugalbanda and the Anzud Bird** | Syro-Mesopotamian | Sumerian | 1 | -2100 |
| Lugalbanda in the Mountain Cave | Syro-Mesopotamian | Sumerian | 1 | -2100 |
| **The lament of Ur** | Syro-Mesopotamian | Sumerian | 1 | -2000 |
| Prophecy of neferti | Egyptian | Old Egyptian hieroglyphs | 2 | -1888 .5 |
| 42 Book of the dead | Egyptian | Old Egyptian hieroglyphs | 2 | -1862 .5 |
| Hymin to the Nile flood | Egyptian | Old Egyptian hieroglyphs | 2 | -1862 .5 |
| Loyalist teaching | Egyptian | Old Egyptian hieroglyphs | 2 | -1858 .5 |
| 21 Book of the dead | Egyptian | Old Egyptian hieroglyphs | 2 | -1852 .5 |
| 38 Book of the dead | Egyptian | Old Egyptian hieroglyphs | 2 | -1852 .5 |
| 77 Book of the dead | Egyptian | Old Egyptian hieroglyphs | 2 | -1852 .5 |
| The instructions of Amenemhe'et | Egyptian | Old Egyptian hieroglyphs | 2 | -1852 .5 |
| Isis and the name of Ra | Egyptian | Old Egyptian hieroglyphs | 2 | -1852 .5 |
| Tale of Sanehat | Egyptian | Hieratic | 1 | -1800 |
| Hammurabi code | Syro-Mesopotamian | Akkadian | 1 | -1754 |
| Enuma elish | Syro-Mesopotamian | Old Babylonian | 1 | -1700 |
| Proverbs I | Syro-Mesopotamian | Sumerian | 2 | -1650 |
| Proverbs II | Syro-Mesopotamian | Sumerian | 2 | -1650 |
| Proverbs III | Syro-Mesopotamian | Sumerian | 2 | -1650 |
| Proverbs IV | Syro-Mesopotamian | Sumerian | 2 | -1650 |
| Proverbs IX | Syro-Mesopotamian | Sumerian | 2 | -1650 |
| Proverbs V | Syro-Mesopotamian | Sumerian | 2 | -1650 |
| Proverbs VII | Syro-Mesopotamian | Sumerian | 2 | -1650 |
| Proverbs VIII | Syro-Mesopotamian | Sumerian | 2 | -1650 |
| Proverbs X | Syro-Mesopotamian | Sumerian | 2 | -1650 |
| Proverbs XI | Syro-Mesopotamian | Sumerian | 2 | -1650 |
| Proverbs XII | Syro-Mesopotamian | Sumerian | 2 | -1650 |
| Proverbs XIII | Syro-Mesopotamian | Sumerian | 2 | -1650 |
| Proverbs XIV | Syro-Mesopotamian | Sumerian | 2 | -1650 |
| Proverbs XIX | Syro-Mesopotamian | Sumerian | 2 | -1650 |
| Proverbs XV | Syro-Mesopotamian | Sumerian | 2 | -1650 |
| Proverbs XVI | Syro-Mesopotamian | Sumerian | 2 | -1650 |
| Proverbs XVII | Syro-Mesopotamian | Sumerian | 2 | -1650 |
| Proverbs XVIII | Syro-Mesopotamian | Sumerian | 2 | -1650 |
| Proverbs XXI | Syro-Mesopotamian | Sumerian | 2 | -1650 |
| Proverbs XXII | Syro-Mesopotamian | Sumerian | 2 | -1650 |
| Proverbs XXIII | Syro-Mesopotamian | Sumerian | 2 | -1650 |
| Proverbs XXIV | Syro-Mesopotamian | Sumerian | 2 | -1650 |
| Proverbs XXV | Syro-Mesopotamian | Sumerian | 2 | -1650 |
| Proverbs XXVI | Syro-Mesopotamian | Sumerian | 2 | -1650 |
| Proverbs XXVIII | Syro-Mesopotamian | Sumerian | 2 | -1650 |
| Proverbs XXVVII | Syro-Mesopotamian | Sumerian | 2 | -1650 |
| The Epic of atrahasis | Syro-Mesopotamian | Akkadian | 2 | -1636 |
| The flood story | Syro-Mesopotamian | Sumerian | 1 | -1600 |
| **26 Book of the dead** | Egyptian | Old Egyptian hieroglyphs | 2 | -1565 |
| 43 Book of the dead | Egyptian | Old Egyptian hieroglyphs | 2 | -1565 |
| **64 Book of the dead** | Egyptian | Old Egyptian hieroglyphs | 2 | -1565 |
| 149 Book of the dead | Egyptian | Old Egyptian hieroglyphs | 2 | -1418 |
| 119 Book of the dead | Egyptian | Old Egyptian hieroglyphs | 2 | -1417 .5 |
| 155 Book of the dead | Egyptian | Old Egyptian hieroglyphs | 2 | -1417 .5 |
| 156 Book of the dead | Egyptian | Old Egyptian hieroglyphs | 2 | -1417 .5 |
| 161 Book of the dead | Egyptian | Old Egyptian hieroglyphs | 2 | -1417 .5 |
| **30 Book of the dead** | Egyptian | Old Egyptian hieroglyphs | 2 | -1417 .5 |
| 7 Book of the dead | Egyptian | Old Egyptian hieroglyphs | 2 | -1417 .5 |
| 83 Book of the dead | Egyptian | Old Egyptian hieroglyphs | 2 | -1417 .5 |
| Adapa and the food of life | Syro-Mesopotamian | Sumerian | 1 | -1400 |
| 100 Book of the dead | Egyptian | Old Egyptian hieroglyphs | 1 | -1400 |
| 101 Book of the dead | Egyptian | Old Egyptian hieroglyphs | 1 | -1400 |
| 102 Book of the dead | Egyptian | Old Egyptian hieroglyphs | 1 | -1400 |
| 112 Book of the dead | Egyptian | Old Egyptian hieroglyphs | 1 | -1400 |
| 115 Book of the dead | Egyptian | Old Egyptian hieroglyphs | 1 | -1400 |
| 124 Book of the dead | Egyptian | Old Egyptian hieroglyphs | 1 | -1400 |
| **1 Book of the dead Papyrus of Nu** | Egyptian | Cursive hieroglyphs | 1 | -1400 |
| 44 Book of the dead | Egyptian | Old Egyptian hieroglyphs | 2 | -1400 |
| 50 Book of the dead | Egyptian | Old Egyptian hieroglyphs | 2 | -1400 |
| 56 Book of the dead | Egyptian | Old Egyptian hieroglyphs | 2 | -1400 |
| 96 Book of the dead | Egyptian | Old Egyptian hieroglyphs | 1 | -1400 |
| 99 Book of the dead | Egyptian | Old Egyptian hieroglyphs | 1 | -1400 |
| **Rigveda** | Hinduist | Indo-european | 2 | -1400 |
| Amarna letter 26 | Syro-Mesopotamian | Akkadian | 2 | -1369 .5 |
| Amarna letter 9 | Syro-Mesopotamian | Akkadian | 2 | -1369 .5 |
| Amarna letter EA12 | Syro-Mesopotamian | Akkadian | 2 | -1369 .5 |
| Amarna letter EA15 | Syro-Mesopotamian | Akkadian | 2 | -1369 .5 |
| Amarna letter EA17 | Syro-Mesopotamian | Akkadian | 2 | -1369 .5 |
| Amarna letter EA23 | Syro-Mesopotamian | Akkadian | 2 | -1369 .5 |
| Amarna letter EA3 | Syro-Mesopotamian | Akkadian | 2 | -1369 .5 |
| Amarna letter EA35 EA38 | Syro-Mesopotamian | Akkadian | 2 | -1369 .5 |
| Amarna letter EA41 | Syro-Mesopotamian | Akkadian | 2 | -1369 .5 |
| Amarna letter EA7 | Syro-Mesopotamian | Akkadian | 2 | -1369 .5 |
| Amarna letter EA8 | Syro-Mesopotamian | Akkadian | 2 | -1369 .5 |
| Amarna letter EA Tushratta to Akhenaten | Syro-Mesopotamian | Akkadian | 2 | -1369 .5 |
| Amarna belief | Egyptian | Old Egyptian hieroglyphs | 2 | -1344 .5 |
| The book of gates | Egyptian | Old Egyptian hieroglyphs | 2 | -1313 .5 |
| Boundary of Stelae | Egyptian | Old Egyptian hieroglyphs | 2 | -1310 |
| 136 Book of the dead | Egyptian | Old Egyptian hieroglyphs | 2 | -1309 .5 |
| 151 Book of the dead | Egyptian | Old Egyptian hieroglyphs | 2 | -1309 .5 |
| 31 Book of the dead | Egyptian | Old Egyptian hieroglyphs | 2 | -1309 .5 |
| 6 Book of the dead | Egyptian | Old Egyptian hieroglyphs | 2 | -1309 .5 |
| 15 Book of the dead | Egyptian | Old Egyptian hieroglyphs | 2 | -1180 .5 |
| 23 Book of the dead | Egyptian | Old Egyptian hieroglyphs | 2 | -1103 .5 |
| Sama veda | Hinduist | Vedic-Sanskrit | 2 | -1100 |
| Yajur veda | Hinduist | Vedic-Sanskrit | 2 | -1100 |
| Atharva veda | Hinduist | Vedic-Sanskrit | 2 | -1100 |
| 17 Book of the dead | Egyptian | Old Egyptian hieroglyphs | 2 | -866 .5 |
| Homer Iliad | Greco-Roman | Ancient Greek | 1 | -800 |
| Homer Odyssey | Greco-Roman | Ancient Greek | 1 | -800 |
| Old Testament 20 Proverbs | Judeo-Christian | Hebrew | 2 | -750 |
| Old Testament 23 Prophet Isaiah | Judeo-Christian | Hebrew | 2 | -750 |
| Old Testament 28 Hosea | Judeo-Christian | Hebrew | 2 | -750 |
| Old Testament 30 Amos | Judeo-Christian | Hebrew | 2 | -750 |
| Old Testament 32 Micah | Judeo-Christian | Hebrew | 2 | -750 |
| Old Testament 7 Judges | Judeo-Christian | Hebrew | 2 | -750 |
| Old Testament 32 Jonah | Judeo-Christian | Hebrew | 2 | -725 |
| Hesiod The Homeric Hymns and Homerica | Greco-Roman | Ancient Greek | 3 | -700 |
| Old Testament 19 Psalms | Judeo-Christian | Hebrew | 2 | -665 .5 |
| **Mahabharata I Adi Parva** | Hinduist | Sanskrit | 2 | -650 |
| **Mahabharata II Sabha Parva** | Hinduist | Sanskrit | 2 | -650 |
| **Mahabharata III Vana Parva** | Hinduist | Sanskrit | 2 | -650 |
| Mahabharata IV Virata Parva | Hinduist | Sanskrit | 2 | -650 |
| Mahabharata V Udyoga Parva | Hinduist | Sanskrit | 2 | -650 |
| Mahabharata VI Bhishma Parva | Hinduist | Sanskrit | 2 | -650 |
| Mahabharata VII Drona Parva | Hinduist | Sanskrit | 2 | -650 |
| Mahabharata VIII Karna Parva | Hinduist | Sanskrit | 2 | -650 |
| Mahabharata IX Shalya Parva | Hinduist | Sanskrit | 2 | -650 |
| Mahabharata X Sauptika Parva | Hinduist | Sanskrit | 2 | -650 |
| Mahabharata XI Stri Parva | Hinduist | Sanskrit | 2 | -650 |
| Mahabharata XII Shanti Parva | Hinduist | Sanskrit | 2 | -650 |
| Mahabharata XIV Ashvamedhika Parva | Hinduist | Sanskrit | 2 | -650 |
| Mahabharata XVI Mausala Parva | Hinduist | Sanskrit | 2 | -650 |
| Mahabharata XVII Mahaprasthanika Parva | Hinduist | Sanskrit | 2 | -650 |
| Mahabharata XIX | Hinduist | Sanskrit | 2 | -650 |
| Old Testament 36 Zephaniah | Judeo-Christian | Hebrew | 2 | -624 .5 |
| Old Testament 35 Habakkuk | Judeo-Christian | Hebrew | 1 | -605 |
| Old Testament 34 Nahum | Judeo-Christian | Hebrew | 1 | -602 |
| Aesop Fables | Greco-Roman | Ancient Greek | 3 | -592 |
| Old Testament 6 Joshua | Judeo-Christian | Hebrew | 2 | -589 .5 |
| Yasna (Avesta) | Persian | Indo European (Avestan) | 3 | -589 .5 |
| **Vendidad (Avesta)** | Persian | Indo European (Avestan) | 3 | -589 .5 |
| **Visperad (Avesta)** | Persian | Indo European (Avestan) | 3 | -589 .5 |
| **Yasht (Avesta)** | Persian | Indo European (Avestan) | 3 | -589 .5 |
| Old Testament 31 Obadiah | Judeo-Christian | Hebrew | 1 | -586 |
| Old Testament 10 Samuel II | Judeo-Christian | Hebrew | 2 | -585 |
| Old Testament 24 Prophet Jeremiah | Judeo-Christian | Hebrew | 2 | -585 |
| Old Testament 25 Lamentations of Jeremiah | Judeo-Christian | Hebrew | 1 | -585 |
| Old Testament 9 Samuel | Judeo-Christian | Hebrew | 2 | -585 |
| Old Testament 26 Prophet Ezekiel | Judeo-Christian | Hebrew | 2 | -582 |
| Old Testament 37 Haggai | Judeo-Christian | Hebrew | 1 | -520 |
| Old Testament 38 Zechariah | Judeo-Christian | Hebrew | 2 | -519 |
| Old Testament 22 Songs of Solomon | Judeo-Christian | Hebrew | 2 | -501 |
| **Old Testament 18 Job** | Judeo-Christian | Hebrew | 2 | -500 |
| Pindar Odes | Greco-Roman | Ancient Greek | 3 | -482 .5 |
| Old Testament 17 Esther | Judeo-Christian | Hebrew | 2 | -475 .5 |
| Bacchylides Dithyramb | Greco-Roman | Latin | 3 | -473 |
| Bacchylides Epinicians | Greco-Roman | Latin | 3 | -473 |
| Herodotus The First Book of the histories called Clio | Greco-Roman | Ancient Greek | 3 | -454 .5 |
| Herodotus An account of Egypt | Greco-Roman | Ancient Greek | 3 | -454 .5 |
| Herodotus The History vol 2 | Greco-Roman | Ancient Greek | 3 | -454 .5 |
| Old Testament 8 Ruth | Judeo-Christian | Hebrew | 2 | -450 |
| Antiphon Anonymous Prosecution for Murder | Greco-Roman | Ancient Greek | 3 | -445 .5 |
| Antiphon On the Choreutes | Greco-Roman | Ancient Greek | 3 | -445 .5 |
| Antiphon On the Murder of Herodes | Greco-Roman | Ancient Greek | 3 | -445 .5 |
| Antiphon Speech 1 | Greco-Roman | Ancient Greek | 3 | -445 .5 |
| Antiphon The Second Tetralogy Prosecution for Accidental Homicide | Greco-Roman | Ancient Greek | 3 | -445 .5 |
| Antiphon The Third Tetralogy Prosecution for Murder Of One Who Pleads Self Defense | Greco-Roman | Ancient Greek | 3 | -445 .5 |
| Old Testament 39 Malachi | Judeo-Christian | Greek | 1 | -445 |
| **Old Testament 1 Genesis** | Judeo-Christian | Hebrew | 2 | -434 |
| Old Testament 2 Exodus | Judeo-Christian | Hebrew | 2 | -434 |
| Old Testament 3 Leviticus | Judeo-Christian | Hebrew | 2 | -434 |
| Old Testament 4 Numbers | Judeo-Christian | Hebrew | 2 | -434 |
| Old Testament 5 Deuteronomy | Judeo-Christian | Hebrew | 2 | -434 |
| Thucydides Stories from Thucydides | Greco-Roman | Ancient Greek | 3 | -430 |
| Thucydides The History of the Peloponnesian War | Greco-Roman | Ancient Greek | 3 | -430 |
| Xenophon Hiero | Greco-Roman | Ancient Greek | 3 | -430 |
| Xenophon On hunting | Greco-Roman | Ancient Greek | 3 | -430 |
| Plato Apology | Greco-Roman | Ancient Greek | 3 | -419 .5 |
| Plato Euthyphro | Greco-Roman | Ancient Greek | 3 | -419 .5 |
| Plato Gorgias | Greco-Roman | Ancient Greek | 3 | -419 .5 |
| Plato Phaedrus | Greco-Roman | Ancient Greek | 3 | -419 .5 |
| Plato Protagoras | Greco-Roman | Ancient Greek | 3 | -419 .5 |
| Plato Republic | Greco-Roman | Ancient Greek | 3 | -419 .5 |
| Plato Sophist | Greco-Roman | Ancient Greek | 3 | -419 .5 |
| Plato Symposium | Greco-Roman | Ancient Greek | 3 | -419 .5 |
| Plato Timaeus | Greco-Roman | Ancient Greek | 3 | -419 .5 |
| Andocides On the Mysteries | Greco-Roman | Ancient Greek | 3 | -415 |
| Andocides Against Alcibiades | Greco-Roman | Ancient Greek | 3 | -415 |
| Andocides On his return | Greco-Roman | Ancient Greek | 3 | -415 |
| Andocides On the peace with Sparta | Greco-Roman | Ancient Greek | 3 | -415 |
| Hippocrates Airs waters places | Greco-Roman | Ancient Greek | 3 | -415 |
| Hippocrates Ancient Medicine | Greco-Roman | Ancient Greek | 3 | -415 |
| Hippocrates Aphorisms | Greco-Roman | Ancient Greek | 3 | -415 |
| Hippocrates On Regimen in Acute Diseases | Greco-Roman | Ancient Greek | 3 | -415 |
| Hippocrates On the epidemics | Greco-Roman | Ancient Greek | 3 | -415 |
| Hippocrates On the sacred disease | Greco-Roman | Ancient Greek | 3 | -415 |
| Lysias Orations | Greco-Roman | Ancient Greek | 3 | -412 .5 |
| Xenophon Anabasis | Greco-Roman | Ancient Greek | 3 | -392 |
| Xenophon Cyropaedia | Greco-Roman | Ancient Greek | 3 | -392 |
| Xenophon Economics | Greco-Roman | Ancient Greek | 3 | -392 |
| Xenophon Hellenica | Greco-Roman | Ancient Greek | 3 | -392 |
| Xenophon On the Art of Horsemanship | Greco-Roman | Ancient Greek | 3 | -392 |
| Xenophon The Memorabilia Recollections of Socrates | Greco-Roman | Ancient Greek | 3 | -392 |
| Xenophon The Symposium | Greco-Roman | Ancient Greek | 3 | -392 |
| Xenophon Ways and means | Greco-Roman | Ancient Greek | 3 | -392 |
| Isocrates Aeropagiticus | Greco-Roman | Ancient Greek | 3 | -387 |
| Isocrates Against the sophists | Greco-Roman | Ancient Greek | 3 | -387 |
| Isocrates Evagoras | Greco-Roman | Ancient Greek | 3 | -387 |
| Isocrates Helen | Greco-Roman | Ancient Greek | 3 | -387 |
| Isocrates Nicocles or Cyprians | Greco-Roman | Ancient Greek | 3 | -387 |
| Isocrates Panegyricus | Greco-Roman | Ancient Greek | 3 | -387 |
| Isocrates Philippus | Greco-Roman | Ancient Greek | 3 | -387 |
| Isocrates To Demonicus | Greco-Roman | Ancient Greek | 3 | -387 |
| Isocrates To Nicocles | Greco-Roman | Ancient Greek | 3 | -387 |
| Isaeus Speeches | Greco-Roman | Ancient Greek | 3 | -384 |
| Horus stelae | Egyptian | Old Egyptian hieroglyphs | 2 | -361 |
| Hyperides Speeches | Greco-Roman | Ancient Greek | 3 | -356 |
| Aristotle Categories | Greco-Roman | Ancient Greek | 3 | -353 |
| Aristotle Ethics | Greco-Roman | Ancient Greek | 3 | -353 |
| Aristotle The Athenian Constitution | Greco-Roman | Ancient Greek | 3 | -353 |
| Aristotle Treatise on Government | Greco-Roman | Ancient Greek | 3 | -353 |
| Demosthenes Exordia | Greco-Roman | Ancient Greek | 3 | -353 |
| Demosthenes Olynthiacs | Greco-Roman | Ancient Greek | 3 | -353 |
| Demosthenes Philippics | Greco-Roman | Ancient Greek | 3 | -353 |
| Aeschines Against Ctesiphon | Greco-Roman | Ancient Greek | 3 | -351 .5 |
| Aeschines Against Timarchus | Greco-Roman | Ancient Greek | 3 | -351 .5 |
| Aeschines On the Embassy | Greco-Roman | Ancient Greek | 3 | -351 .5 |
| The Upanishads | Hinduist | Sanskrit | 2 | -350 |
| Old Testament 29 Joel | Judeo-Christian | Hebrew | 2 | -350 |
| Demades On the twelve years | Greco-Roman | Ancient Greek | 3 | -349 |
| Dinarchus Against Aristogiton | Greco-Roman | Ancient Greek | 3 | -326 |
| Dinarchus Against Demosthenes | Greco-Roman | Ancient Greek | 3 | -326 |
| Dinarchus Against Philocles | Greco-Roman | Ancient Greek | 3 | -326 |
| Old Testament 11 Kings | Judeo-Christian | Hebrew | 2 | -325 |
| Old Testament 12 Kings II | Judeo-Christian | Hebrew | 2 | -325 |
| Old Testament 13 Chronicles | Judeo-Christian | Hebrew | 2 | -325 |
| Old Testament 14 Chronicles II | Judeo-Christian | Hebrew | 2 | -325 |
| Old Testament 21 Ecclesiastes | Judeo-Christian | Hebrew | 2 | -315 |
| Epicurus Letter to Menoceus | Greco-Roman | Ancient Greek | 3 | -305 .5 |
| Epicurus The doctrines | Greco-Roman | Ancient Greek | 3 | -305 .5 |
| The Burden of Isis | Egyptian | Old Egyptian hieroglyphs | 1 | -300 |
| Apollonius Argonautica | Greco-Roman | Ancient Greek | 3 | -250 |
| Apollonius Argonautica Book 2 | Greco-Roman | Ancient Greek | 3 | -250 |
| Apollonius Argonautica Book 3 | Greco-Roman | Ancient Greek | 3 | -250 |
| Apollonius Argonautica Book 4 | Greco-Roman | Ancient Greek | 3 | -250 |
| Old Testament 15 Ezra | Judeo-Christian | Hebrew | 2 | -200 |
| Rosetta Stone | Egyptian | Old Egyptian hieroglyphs; Demotic; Ancient greek | 1 | -196 |
| The story of the book of thot | Egyptian | Old Egyptian hieroglyphs | 2 | -167 .5 |
| Old Testament 27 Daniel | Judeo-Christian | Hebrew | 1 | -164 |
| Old Testament 16 Nehemiah | Judeo-Christian | Hebrew | 2 | -150 .5 |
| Apollodorus Epit | Greco-Roman | Ancient Greek | 3 | -150 |
| Apollodorus Library and Epitome Book 1 | Greco-Roman | Ancient Greek | 3 | -150 |
| Apollodorus Library and Epitome Book 2 | Greco-Roman | Ancient Greek | 3 | -150 |
| Cicero Letters | Greco-Roman | Latin | 3 | -74 .5 |
| Cicero Treatise on the Laws Book 1 | Greco-Roman | Latin | 3 | -74 .5 |
| Cicero Treatise on the Laws Book 3 | Greco-Roman | Latin | 3 | -74 .5 |
| Cicero Treatise on the law books 2 | Greco-Roman | Latin | 3 | -74 .5 |
| **Julius Caesar Commentaries of the Civil War** | Greco-Roman | Latin | 3 | -72 |
| Julius Caesar Supplement of Dionysius vossius to Caesar's first book of the civil war | Greco-Roman | Latin | 3 | -72 |
| Julius Caesar The Gallic Wars | Greco-Roman | Latin | 3 | -72 |
| Hirtius The African Wars | Greco-Roman | Latin | 3 | -66 .5 |
| Hirtius The Alexandrian Wars | Greco-Roman | Latin | 3 | -66 .5 |
| Hirtius The Spanish Wars | Greco-Roman | Latin | 3 | -66 .5 |
| Diodorus Fragments of Book 10 | Greco-Roman | Ancient Greek | 2 | -60 |
| Diodorus Fragments of Book 9 | Greco-Roman | Ancient Greek | 2 | -60 |
| Horace Odes | Greco-Roman | Latin | 3 | -46 |
| Augustus The deeds | Greco-Roman | Latin | 3 | -24 .5 |
| Livy The History of Rome 1 | Greco-Roman | Latin | 3 | -23 .5 |
| Livy The History of Rome 2 | Greco-Roman | Latin | 3 | -23 .5 |
| Livy The History of Rome 3 | Greco-Roman | Latin | 3 | -23 .5 |
| Strabo Geography | Greco-Roman | Ancient Greek | 3 | -20 |
| Laws of Manu | Hinduist | Sanskrit | 2 | 0 |
| New Testament 13 First Epistle of Paul to Thessalonians | Judeo-Christian | Greek | 1 | 51 |
| New Testament 11 Epistle of Paul to Philipians | Judeo-Christian | Greek | 2 | 54 .5 |
| New Testament 19 Epistle of Paul to Philemon | Judeo-Christian | Greek | 2 | 54 .5 |
| New Testament 9 Epistle of Paul to Galatians | Judeo-Christian | Greek | 1 | 55 |
| **New Testament 7 First Epistle of Paul to Corinthians** | Judeo-Christian | Greek | 1 | 56 |
| New Testament 8 Second Epistle of Paul to Corinthians | Judeo-Christian | Greek | 1 | 56 |
| New Testament 6 Epistle of Paul to Romans | Judeo-Christian | Greek | 1 | 57 |
| New Testament 15 Second Epistle of Paul to Thessalonians | Judeo-Christian | Greek | 1 | 60 |
| New Testament 12 Epistle of Paul to Colossians | Judeo-Christian | Greek | 2 | 66 |
| Josephus Against Apion | Greco-Roman | Ancient Greek | 3 | 68 .5 |
| Josephus Antiquities of the Jews | Greco-Roman | Ancient Greek | 3 | 68 .5 |
| Josephus Autobiography | Greco-Roman | Ancient Greek | 3 | 68 .5 |
| Josephus Jewish Wars | Greco-Roman | Ancient Greek | 3 | 68 .5 |
| New Testament 2 Gospel According to St Mark | Judeo-Christian | Greek | 2 | 69 |
| New Testament 21 General Epistle of James | Judeo-Christian | Greek | 2 | 75 |
| New Testament 27 Epistle of Jude | Judeo-Christian | Greek | 2 | 78 |
| New Testament 22 First Epistle of Peter | Judeo-Christian | Greek | 2 | 82 .5 |
| Plutarch Alcibiades | Greco-Roman | Ancient Greek | 3 | 83 |
| Plutarch Aristides | Greco-Roman | Ancient Greek | 3 | 83 |
| Plutarch Cimos | Greco-Roman | Ancient Greek | 3 | 83 |
| Plutarch Lysander | Greco-Roman | Ancient Greek | 3 | 83 |
| Plutarch Nicias | Greco-Roman | Ancient Greek | 3 | 83 |
| Plutarch Pericles | Greco-Roman | Ancient Greek | 3 | 83 |
| Plutarch Solon | Greco-Roman | Ancient Greek | 3 | 83 |
| Plutarch Themistocles | Greco-Roman | Ancient Greek | 3 | 83 |
| Plutarch Theseus | Greco-Roman | Ancient Greek | 3 | 83 |
| New Testament 10 Epistle of Paul to Ephesians | Judeo-Christian | Greek | 2 | 85 |
| New Testament 1 Gospel according to Matthew | Judeo-Christian | Greek | 1 | 85 |
| New Testament 20 Epistle of Paul to Hebrews | Judeo-Christian | Greek | 2 | 85 |
| **New Testament 3 Gospel According to St Luke** | Judeo-Christian | Greek | 2 | 85 |
| Tacitus A Dialogue Concerning Oratory Or The Causes Of Corrupt Eloquence | Greco-Roman | Latin | 3 | 86 .5 |
| Tacitus the Rein of Tiberius | Greco-Roman | Latin | 3 | 86 .5 |
| **Tacitus A treatise on the situation manners and inhabitants of Germany** | Greco-Roman | Latin | 3 | 86 .5 |
| New Testament 28 Revelations | Judeo-Christian | Greek | 1 | 95 |
| Epictetus Manual of Epictetus | Greco-Roman | Ancient Greek | 3 | 95 |
| Epictetus The Discourses | Greco-Roman | Ancient Greek | 3 | 95 |
| New Testament 5 Acts of Apostles | Judeo-Christian | Greek | 1 | 97 .5 |
| Vishnu Purana | Hinduist | Sanskrit | 2 | 100 |
| New Testament 16 First Epistle of Paul to Timothy | Judeo-Christian | Greek | 1 | 100 |
| New Testament 17 Second Epistle of Paul to Timothy | Judeo-Christian | Greek | 1 | 100 |
| New Testament 18 Epistle of Paul to Titus | Judeo-Christian | Greek | 1 | 100 |
| New Testament 24 First Epistle of John | Judeo-Christian | Greek | 1 | 100 |
| New Testament 25 Second Epistle of John | Judeo-Christian | Greek | 1 | 100 |
| New Testament 26 Third Epistle of John | Judeo-Christian | Greek | 1 | 100 |
| New Testament 4 Gospel According to St John | Judeo-Christian | Greek | 1 | 100 |
| New Testament 23 Second Epistle of Peter | Judeo-Christian | Greek | 1 | 110 |
| Pausanias Description of Greece | Greco-Roman | Ancient Greek | 3 | 145 |
| **Apuleius Apologia** | Greco-Roman | Latin | 3 | 147 |
| Apuleius The Golden Asse | Greco-Roman | Latin | 3 | 147 |
| Apuleius The Florida | Greco-Roman | Latin | 3 | 147 |
| Marcus Aurelius Thoughts | Greco-Roman | Latin | 3 | 150 .5 |
| Marcus Aureliu Memories | Greco-Roman | Latin | 3 | 150 .5 |
| Galen On the natural faculties | Greco-Roman | Ancient Greek | 3 | 164 .5 |
| Talmud 1 | Judeo-Christian | Aramaic | 2 | 251 |
| Talmud 2 | Judeo-Christian | Aramaic | 2 | 251 |
| Talmud 3 | Judeo-Christian | Aramaic | 2 | 251 |
| Talmud 4 | Judeo-Christian | Aramaic | 2 | 251 |
| Talmud 5 | Judeo-Christian | Aramaic | 2 | 251 |
| Talmud 6 | Judeo-Christian | Aramaic | 2 | 251 |
| Talmud 7 | Judeo-Christian | Aramaic | 2 | 251 |
| Talmud 8 | Judeo-Christian | Aramaic | 2 | 251 |
| Talmud 9 | Judeo-Christian | Aramaic | 2 | 251 |
| Sikand Gûmânîk Vigâr | Persian | Middle-Persian (Pahlavi) | 2 | 251 |
| Pahlavi Sad Dar | Persian | Middle-Persian (Pahlavi) | 2 | 251 |
| Porphyry Against the Christians | Greco-Roman | Ancient Greek | 3 | 269 .5 |
| Porphyry Introduction to categories of logic | Greco-Roman | Ancient Greek | 3 | 269 .5 |
| Porphyry Letter to Marcella | Greco-Roman | Ancient Greek | 3 | 269 .5 |
| Porphyry On abstinence from animal food | Greco-Roman | Ancient Greek | 3 | 269 .5 |
| Porphyry On the cave of Nymphus | Greco-Roman | Ancient Greek | 3 | 269 .5 |
| Porphyry On the faculties of the soul | Greco-Roman | Ancient Greek | 3 | 269 .5 |
| Porphyry The life of Plotinus | Greco-Roman | Ancient Greek | 3 | 269 .5 |
| Porphyry The life of Pythagoras | Greco-Roman | Ancient Greek | 3 | 269 .5 |
| Porphyry to the Prophet Anebo greeting. | Greco-Roman | Ancient Greek | 3 | 269 .5 |
| Pahlavi Bundahis | Persian | Middle-Persian (Pahlavi) | 2 | 301 |
| Pahlavi Zâd Spararam | Persian | Middle-Persian (Pahlavi) | 2 | 301 |
| Pahlavi Bahman Yasts | Persian | Middle-Persian (Pahlavi) | 2 | 301 |
| Pahlavi Shayâst la shayâst | Persian | Middle-Persian (Pahlavi) | 2 | 301 |
| Pahlavi Dâdistân Î Dînîk | Persian | Middle-Persian (Pahlavi) | 2 | 301 |
| Epistle of Mânushihar | Persian | Middle-Persian (Pahlavi) | 2 | 301 |
| Epistle of Mânushihar II | Persian | Middle-Persian (Pahlavi) | 2 | 301 |
| Matsya Purana | Hinduist | Sanskrit | 2 | 350 |
| Homily III John Chrisostome | Medieval | Greek | 2 | 386 .5 |
| Homily II John Chrisostome | Medieval | Grreek | 2 | 386 .5 |
| Homily I John Chrisostome | Medieval | Grreek | 2 | 386 .5 |
| **St Augustine Confessions** | Judeo-Christian | Latin | 2 | 397 |
| Markandeya Purana | Hinduist | Sanskrit | 2 | 400 |
| Vayu Purana | Hinduist | Sanskrit | 2 | 400 |
| **City of God St Augustine** | Medieval | Latin | 1 | 426 |
| Book of Arda Viraf | Persian | Middle-Persian (Pahlavi) | 2 | 437 .5 |
| **Regulae Pastoralis Pope Gregory I** | Medieval | Latin | 1 | 590 |
| Skanda Purana | Hinduist | Sanskrit | 2 | 600 |
| **The Noble Qu'ran** | Medieval | Arabic | 2 | 620 .5 |
| Linga Purana | Hinduist | Sanskrit | 2 | 750 |
| 1001 Arabian Nights | Medieval | Arabic | 1 | 750 .5 |
| Dinkard book 5 | Persian | Middle-Persian (Pahlavi) | 2 | 851 |
| Dinkard book 6 | Persian | Middle-Persian (Pahlavi) | 2 | 851 |
| Dînâ Î Maînôg Î | Persian | Middle-Persian (Pahlavi) | 2 | 851 |
| Dinkard book 8 | Persian | Middle-Persian (Pahlavi) | 2 | 851 |
| Dinkard book 9 | Persian | Middle-Persian (Pahlavi) | 2 | 851 |
| Padma Purana | Hinduist | Sanskrit | 2 | 875 |
| Agni Purana | Hinduist | Sanskrit | 2 | 900 |
| Bhagavat Purana | Hinduist | Sanskrit | 2 | 900 |
| Garuda Purana | Hinduist | Sanskrit | 2 | 900 |
| Codex Junius 11 | Medieval | Old English | 2 | 965 |
| Story of Beowulf | Medieval | Old English | 1 | 992 |
| Vamana Purana | Hinduist | Sanskrit | 2 | 1000 |
| Book of Healing Avicenna | Medieval | Persian | 2 | 1023 .5 |
| Vahara Purana | Hinduist | Sanskrit | 2 | 1100 |
| Geoffrey Mounmouth Historia Regum Britanniae | Medieval | Latin | 1 | 1136 |
| Alexiada Anna Comnena | Medieval | Attic Greek | 1 | 1148 |
| Geoffrey of Monmouth Vita Merlini | Medieval | Latin | 1 | 1150 |
| Shiva Maha Purana | Hinduist | Sanskrit | 2 | 1200 |
| Thomas of England Tristan and Isolde | Medieval | Celtic | 1 | 1227 |
| Summa Theologica I | Medieval | Latin | 1 | 1269 .5 |
| Summa Theologica II II | Medieval | Latin | 1 | 1269 .5 |
| Narada Purana | Hinduist | Sanskrit | 2 | 1300 |
| Bhrama Purana | Hinduist | Sanskrit | 2 | 1350 |
| Secretum Petrarca | Medieval | Latin | 2 | 1350 |
| **Decameron Giovanni Boccacio** | Medieval | Latin | 2 | 1350 .5 |
| Catherine of Siena Letters | Medieval | Tuscan | 1 | 1375 |
| Christine du Pisan Book of the Duke | Medieval | French | 1 | 1431 |
| Tales of King Arthur | Modern | Middle English | 1 | 1485 |
| The White Knight | Modern | Valencian | 1 | 1490 |
| Rotterdan Praise of Folly | Modern | Latin | 1 | 1511 |
| Martin Luther On the freedom of a christian | Modern | Latin | 1 | 1520 |
| Machiavelli The Prince | Modern | Italian | 1 | 1532 |
| John Calvin Institutes of Christian Religion | Modern | French | 1 | 1541 |
| Rabelais Gargantua and Pantagruel | Modern | French | 1 | 1548 |
| Teresa of Avila The life of Teresa of Jesus | Modern | Spanish | 1 | 1567 |
| Cervantes Don Quijote | Modern | Early Modern Spanish | 2 | 1610 |
| Descartes Discourse on the method | Modern | French | 1 | 1637 |
| Aphra Behn Ten pleasures of marriage | Modern | Early Modern English | 1 | 1682 |
| Locke Second treatise on government | Modern | French | 1 | 1689 |
| Daniel Defoe Robinson Crusoe | Modern | English | 1 | 1719 |
| Voltaire Zadig | Modern | French | 1 | 1747 |
| Voltaire Candide | Modern | French | 1 | 1759 |
| Emile Jean Jacques Russeau | Modern | French | 1 | 1762 |
| Diderot This is not a story | Modern | French | 1 | 1772 |
| Goethe The Sorrows of Young Wherther | Modern | German | 1 | 1774 |
| Immanuel Kant Critique of pure reason | Modern | German | 1 | 1781 |
| Mary Wollstonecraft Rights of woman | Modern | English | 1 | 1792 |
| Jane Austen Pride and Prejudice | Contemporary | English | 1 | 1813 |
| Mary Shelley Frankenstein | Contemporary | English | 1 | 1818 |
| Honoré de Balzac The human comedy | Contemporary | French | 2 | 1824 .5 |
| Stendhal The red and the black | Contemporary | French | 1 | 1830 |
| Edgar Allan Poe Complete Tales | Contemporary | English | 3 | 1840 .5 |
| Charlotte Brontë Jane Eyre | Contemporary | English | 1 | 1847 |
| Fyodor Dostoyevski Crime and Punishment | Contemporary | Russian | 1 | 1866 |
| Oscar Wilde Picture of Dorian Gray | Contemporary | English | 1 | 1890 |
| F. Scott Fitzgerald The great Gatsby | Contemporary | English | 1 | 1925 |
| Virginia Woolf Mrs Dalloway | Contemporary | English | 1 | 1925 |
| William Faulkner The sound and the fury | Contemporary | English | 1 | 1929 |
| William Faulkner As I lay dying | Contemporary | English | 1 | 1930 |
| John Steinbeck Of mice and men | Contemporary | English | 1 | 1937 |
| John Steinbeck Grapes of Wrath | Contemporary | English | 1 | 1939 |
| Jorge Luis Borges Fictions | Contemporary | Spanish | 1 | 1944 |
| Isaac Asimov I, Robot | Contemporary | English | 1 | 1950 |
| J. D. Salinger The catcher in the rye | Contemporary | English | 1 | 1951 |
| Ernest Hemingway The old man and the sea | Contemporary | English | 1 | 1952 |
| Guimaraes Rosa The devil to pay in backlands | Contemporary | Portuguese | 1 | 1956 |
| Gabriel Garcia Marquez 100 years of solitude | Contemporary | Spanish | 1 | 1967 |
| Clarice Lispector Hour of the star | Contemporary | Portuguese | 1 | 1977 |
| Gore Vidal Creation | Contemporary | English | 1 | 1981 |
| Haruki Murakami Norwegian Wood | Contemporary | Japanese | 1 | 1987 |
| John Krakauer Into the Wild | Contemporary | English | 1 | 1996 |
| Neil Gaiman Stardust | Contemporary | English | 1 | 1999 |
| Neil Gaiman American Gods | Contemporary | English | 1 | 2001 |
| Liam Callanan Cloud Atlas | Contemporary | English | 1 | 2004 |
| Markus Zuzak The book thief | Contemporary | English | 1 | 2005 |
| J. K. Rowling Harry Potter and the Deathly Hallows | Contemporary | English | 1 | 2007 |
| Emma Donoghue Room | Contemporary | English | 1 | 2010 |
| Michel Laub Diary of the fall | Contemporary | Portuguese | 1 | 2011 |

**Supplementary Table 3: Statistically significant differences among ontogenetic and historical datasets.** Significant p values indicated in bold (Bonferroni correction for 40 comparisons, alpha = 0.00125).

| KW and p values of post-hocs tests | Nodes | RE | LSC | ASP |
|---|---|---|---|---|
| **Kruskal-Wallis** | **7.48e−038** | **3.95e−037** | **3.59e−065** | **3.97e−037** |
| **Healthy <12 yo x Healthy >12 yo** | **0.0000** | 0.0079 | **0.0000** | **0.0000** |
| **Healthy <12 yo x Psychosis** | 0.6992 | 0.9077 | 0.3311 | 0.0156 |
| **Healthy >12 yo x Psychosis** | **0.0000** | 0.0074 | **0.0002** | **0.0007** |
| **Healthy <12 yo x PreAxial** | 0.0988 | 0.9343 | 0.7987 | 0.0138 |
| **Healthy <12 yo x PostAxial** | **0.0000** | **0.0000** | **0.0000** | **0.0000** |
| **Healthy >12 yo x PreAxial** | **0.0000** | **0.0008** | **0.0000** | **0.0000** |
| **Healthy >12 yo x PostAxial** | 0.0028 | **0.0000** | **0.0000** | 0.2648 |
| **Psychosis x PreAxial** | 0.0628 | 0.8338 | 0.0932 | 0.6479 |
| **Psychosis x PostAxial** | **0.0000** | **0.0000** | **0.0000** | **0.0000** |
| **PreAxial x PostAxial** | **0.0000** | **0.0000** | **0.0000** | **0.0000** |

**Supplementary Table 4: Spearman correlations between graph attributes and years of age.** Significant p values indicated in bold (Bonferroni correction for 8 comparisons (2 groups * 4 attributes), alpha = 0.0063).

| | Spearman Correlation | Nodes | RE | LSC | ASP |
|---|---|---|---|---|---|
| **Control** | **Rho** | 0.36 | -0.22 | 0.40 | 0.41 |
| | **p value** | **0.0000** | 0.0118 | **0.0000** | **0.0000** |
| **Psychosis** | **Rho** | -0.02 | -0.04 | 0.17 | 0.06 |
| | **p value** | 0.8919 | 0.7744 | 0.1806 | 0.6178 |

**Supplementary Table 5: Spearman correlations between graph attributes and years of education.** Significant p values indicated in bold. (Bonferroni correction for 8 comparisons (2 groups * 4 attributes), alpha = 0.0063).

| | Spearman Correlation | Nodes | RE | LSC | ASP |
|---|---|---|---|---|---|
| **Control** | **Rho** | 0.49 | -0.33 | 0.45 | 0.51 |
| | **p value** | **0.0000** | **0.0001** | **0.0000** | **0.0000** |
| **Psychosis** | **Rho** | 0.06 | -0.01 | 0.19 | 0.17 |
| | **p value** | 0.6578 | 0.9253 | 0.1294 | 0.1750 |

**Supplementary Table 6: Correlations of graph attributes with a multiple linear combination of education and age confirm the predominance of the former.** Significant p values indicated in bold (Bonferroni correction for 8 comparisons (2 groups * 4 attributes), alpha = 0.0063).

| | Nodes | RE | LSC | ASP |
|---|---|---|---|---|
| **$R^2$** | 0.16 | 0.09 | 0.23 | 0.26 |
| **p** | **0.0000** | **0.0025** | **0.0000** | **0.0000** |
| **Coef AGE** | -0.0067 | 0.0023 | 0.0578 | 0.0050 |
| **Coef EDU** | 0.1195 | -0.0500 | 0.2353 | 0.0394 |
| **Coef EDU - Coef AGE** | 0.1128 | 0.0478 | 0.1776 | 0.0344 |

**Supplementary Table 7: Statistically significant differences among ontogenetic and historical datasets.** Significant Spearman correlations indicated in bold.

| For years of education | Goodness of Fit | Nodes | RE | LSC | ASP |
|---|---|---|---|---|---|
| **Control** | **R Square** | **0.85** | **0.95** | **0.83** | **0.52** |
| | **SSE** | **7.81** | **0.63** | **45.58** | **0.36** |
| | **RMSE** | **0.53** | **0.15** | **1.28** | **0.11** |
| | **a** | **24.56** | **1.07** | **18.68** | **4.94** |
| | Τ | **0.63** | **0.28** | **13.34** | **11.06** |
| | c | 19.43 | 4.33 | 8.32 | 3.85 |
| | \|a-c\| | 5.13 | 3.26 | 10.36 | 1.08 |
| **Psychosis** | **R Square** | 0.01 | 0.01 | 0.42 | 0.05 |
| | **SSE** | 9.16 | 1.96 | 137.30 | 1.33 |
| | **RMSE** | 0.53 | 0.24 | 2.04 | 0.20 |
| | **a** | 22.53 | 1.55 | 18.84 | 4.43 |
| | Τ | 29.99 | 1.12 | 14.94 | 3.71 |
| | c | 23.48 | 0.00 | 6.69 | 3.59 |
| | \|a-c\| | 0.95 | 1.55 | 12.15 | 0.85 |

**Supplementary Table 8: Temporally randomized ontogenetic data (1,000 surrogations) do not correlate with graph attributes.**

| Ontogenesis | Nodes | RE | LSC | ASP |
|---|---|---|---|---|
| Count (smaller p than real data) | 0 | 0 | 0 | 0 |
| Count (higher \|Rho\| than real data) | 0 | 0 | 0 | 0 |
| Count (higher \|$R^2$\| than real data) | 0 | 4 | 0 | 0 |

**Supplementary Table 9: For historical data, parameters for Spearman and exponential correlations of graph attributes with time.** Significant correlations indicated in bold (Bonferroni correction for 4 comparisons, alpha = 0.0125).

| Spearman | Nodes | RE | LSC | ASP |
|---|---|---|---|---|
| **Rho** | 0.52 | -0.49 | 0.49 | 0.54 |
| **p** | **4.23E-33** | **4.56E-29** | **7.82E-28** | **5.92E-35** |
| **Goodness** | **Nodes** | **RE** | **LSC** | **ASP** |
| **R square** | 0.27 | 0.28 | 0.38 | 0.30 |
| **R adjusted** | 0.26 | 0.27 | 0.38 | 0.30 |
| **SSE** | 543.51 | 119.61 | 3411.53 | 69.64 |
| **RMSE** | 1.11 | 0.52 | 2.77 | 0.40 |
| **Asymptotic a** | 30.00 | 0.03 | 19.15 | 29.00 |
| **Characteristic time** | 5120 | -1028 | -1503 | 98873 |
| **Coefficient c** | 22.31 | 2.59 | 1.00 | 3.69 |

**Supplementary Table 10: Temporally randomized historical data (1,000 surrogations) do not correlate with graph attributes.**

| History | Nodes | RE | LSC | ASP |
|---|---|---|---|---|
| Count (smaller p than real data) | 0 | 0 | 0 | 0 |
| Count (higher \|Rho\| than real data) | 0 | 0 | 0 | 0 |
| Count (higher $\|R^2\|$ than real data) | 0 | 0 | 0 | 0 |

## Supplementary References